\documentclass[aps,prb,preprint,groupedaddress]{revtex4}
\usepackage{graphicx}

\begin{document}


\title{MnAs dots grown on GaN(000$\overline 1$)-(1x1) surface}

\author{I.A. Kowalik}
\author{B.J. Kowalski}
\author{R.J. Iwanowski}
\author{K. Kopalko}
\author{E. {\L}usakowska}
\author{M. Sawicki}

\affiliation{Institute of Physics, Polish Academy of Sciences, Aleja Lotnik\'ow 32/46,
PL-02 668 Warsaw, Poland}
\author{J. Sadowski}
\affiliation{MAX-lab, Lund University, SE-221 00 Lund, Sweden\\ and Institute of Physics, Polish Academy of Sciences, Aleja Lotnik\'ow 32/46, PL-02 668 Warsaw, Poland}
\author{M. Adell}
\affiliation{Department of Physics, Chalmers University of Technology and G\"{o}teborg
University, SE-412 96 G\"{o}teborg, Sweden}
\author{I. Grzegory, S. Porowski}
\affiliation{Institute of High Pressure Physics, Polish Academy of Sciences, Soko{\l}owska 29, PL-01 141 Warsaw, Poland}

\date{\today}

\begin{abstract}
MnAs has been grown by means of MBE on the GaN(000$\overline 1$)-(1x1) surface. Two options of initiating the crystal growth were applied: (a) a regular MBE procedure (manganese and arsenic were delivered simultaneously) and (b) subsequent deposition of manganese and arsenic layers. It was shown that spontaneous formation of MnAs dots with the surface density of 1$\cdot 10^{11}$ cm$^{-2}$ and $2.5\cdot 10^{11}$ cm$^{-2}$, respectively (as observed by AFM), occurred for the layer thickness higher than 5 ML. Electronic structure of the MnAs/GaN systems was studied by resonant photoemission spectroscopy. That led to determination of the Mn 3d - related contribution to the total density of states (DOS) distribution of MnAs. It has been proven that the electronic structures of the MnAs dots grown by the two procedures differ markedly. One corresponds to metallic, ferromagnetic NiAs-type MnAs, the other is similar to that reported for half-metallic zinc-blende MnAs. Both system behave superparamagnetically (as revealed by magnetization measurements), but with both the blocking temperatures and the intra-dot Curie temperatures substantially different. The intra-dot Curie temperature is about 260~K for the former system while markedly higher than room temperature for the latter one. Relations between growth process, electronic structure and other properties of the studied systems are discussed. Possible mechanisms of half-metallic MnAs formation on GaN are considered.
\end{abstract}


\maketitle

\section{Introduction}
Manganese arsenide has been subjected to investigations for a century, at least, since the commonly cited paper by Heusler \cite{heusler04} were published. Its coupled structural and magnetic properties attracted particular interest. At room temperature, it exists in the ferromagnetic $\alpha$ phase, with hexagonal NiAs structure. It transforms into the orthorombic, paramagnetic $\beta$ phase by a first order phase transition at 40$^0$C.\cite{deblois63} At 130$^0$C, the crystalline structure of MnAs changes again to hexagonal (NiAs-like) one and the paramagnetic $\gamma$ phase is formed.\cite{wills54} Interesting magnetic properties of MnAs and other manganese pnictides inspired extensive theoretical studies aimed at determination of the electronic structure and the nature of magnetism of these compounds.\cite{goodenough63,albers64,barner77,%
sandratskii81,podloucky84jmm,podloucky84ssc,coehoorn85,motizuki87,shirai98,ravindran99} The results, together with the experimental data related to the band structure,\cite{bouwma73,liang77,chen76} showed that the electronic structure of MnAs and related compounds should be described in the model of itinerant d-electrons with strong As 4p -- Mn 3d hybridization. Due to that the density of electronic states at the Fermi level is relatively low in these materials. The maxima related to the Mn 3d states in ferromagnetic MnSb occur at about 2.5 eV above and 1 eV below the Fermi level.\cite{coehoorn85} Such description of the electronic structure of manganese pnictides accounts for experimental results concerning their magnetic properties and the specific heat.

Recently, MnAs and related compounds regained considerable interest in the context of emerging projects of spin--electronic (spintronic) devices. Ferromagnetic materials compatible with contemporary semiconductor technology would be particularly important for such applications. \cite{prinz90} Ferromagnetic layers would be indispensable in structures designed as emitters, controllers or detectors of spin-polarized currents. Ferromagnetic materials can also be applied in formation of nano-magnets as sources of local magnetic fields in spin-- or magneto--electronic systems. Among ferromagnetic materials suitable for such purposes, selected compounds of Mn (like GaMn and MnAs) are considered. \cite{tanaka93,kioseoglu02,akeura95,tanaka99,ramsteiner02} Possibility of spontaneous formation of nano-structures or engineering of the layer properties by suitable selection of substrate structure and parameters would lead to even more sophisticated applications, like arrays of nano-magnets or systems of coupled ferromagnetic dots. In this context, MnAs is especially interesting, due to the coupling between its magnetic properties and structure. \cite{ney04} This inspired attempts to perform overgrowth of MnAs on various substrates (e.g. Si(001) \cite{akeura95}, GaAs(001) \cite{tanaka94,ney04}, GaAs(111) \cite{sadowski00}, GaMnAs\cite{sadowski05apl}, ZnSe \cite{berry00}). MnAs-based systems became important examples of so called heteroepitaxy of dissimilar materials \cite{ploog01,trampert02}. It is a technique exploiting large lattice misfit and strong strain at the interface in order to design the structure and adequate properties of the epilayer. On the other side, possibility of modification of the properties of MnAs induced by a choice of the substrate and deposition mode expands the range of MnAs-based systems supposed of exhibiting ferromagnetic properties. This is also important from the point of view of basic research and may lead to better understanding of mechanisms leading to magnetic interactions, previously investigated for bulk MnAs.

In this work, we have performed complex studies of the MnAs epilayers grown (by MBE) on GaN(000$\overline 1$)-(1x1) surface, which included investigations of an influence of the growth conditions on their morphology, magnetic properties and electronic structure. Both materials have hexagonal structure in the plane perpendicular to the c axis, where $a_{MnAs} > a_{GaN}$. Due to this misfit, spontaneous formation of MnAs dots in this system can be expected (contrary to the case of MnAs/GaAs(111)). In order to verify this hypothesis, we have deposited MnAs epilayers on the (000$\overline 1$) GaN surface under continuous {\em in situ} monitoring of their thickness and crystalline structure (by electron diffraction (RHEED)). This, together with {\em ex situ} atomic force microscopy (AFM) results, confirmed that islands of MnAs are formed for the layers thicker that 5 ML. We apply two methods of growth initiation - both of them lead to dots formation. However, electronic structures probed by means of resonant photoemission spectroscopy turns out to be different. The magnetization measurements prove that the ensembles of dots exhibit superparamagnetic properties, although the blocking temperatures and the Curie temperatures characterizing these two systems differ markedly. We present analysis and comparison of the data acquired for samples grown by regular MBE procedure (some of them preliminarily presented in Ref.~\onlinecite{kowalski04app}) and by the process initiated by subsequent deposition of manganese and arsenic. The use of resonant photoemission spectroscopy enables us to determine the electronic states distribution of MnAs layer and, especially, a partial contribution provided by the Mn 3d electrons. The changes revealed in the electronic structure of MnAs dots and their relation to growth procedure and other properties of the investigated systems are discussed.

\section{Experimental details}
The growth of MnAs layers and their electronic structure were investigated in the National Electron Accelerator Laboratory for Nuclear Physics and Synchrotron Radiation Research (MAX-lab), Lund University, Sweden. The layers were grown on bulk GaN substrates by means of MBE technique. The GaN crystals with hexagonal crystalline structure were grown by means of a high pressure technique at the Institute of High Pressure Physics, Polish Academy of Sciences, Warsaw, Poland. The (000$\overline 1$) surfaces of the GaN substrates used in our experiments were initially prepared by {\em ex situ} mechano -- chemical polishing. Prior to the MnAs growth and photoemission measurements, they were introduced into the UHV system and subjected to an {\em in situ} cleaning procedure consisting of the cycles of Ar$^+$ ion bombardment and subsequent annealing at $500^0$ C. As previously shown \cite{kowalski04ss}, such a procedure leads to a clean and well-ordered GaN (000$\overline 1$)-(1x1) surface. In the present experiments, the surface crystallinity was assessed {\em in situ} by electron diffraction observations (LEED and RHEED).

The layers of MnAs were grown stepwise in a KRYOVAK III-V MBE system directly attached to the spectrometer. The electronic structure of the system was investigated {\em in situ} at each stage of the growth process by means of photoelectron spectroscopy. The layer structure was assessed by RHEED (during growth in the MBE system) and LEED (in the photoelectron spectrometer) techniques. The morphology of the MnAs layers with dots formed on the surface was studied {\em ex situ} by AFM. The magnetic properties of this system were determined by the superconducting quantum interference device (SQUID) technique.

The photoemission experiments were carried out at the beamline 41 of MAX-lab. The overall energy resolution was kept around 150 meV, and the angular resolution was about 2$^0$. The origin of the binding energy scale was set at the Fermi level (determined for the reference metal sample). The angle between incoming photon beam and the normal to the surface was kept at 45$^0$. The spectra were normalized to the monochromator output and photon flux variations.

\section{Results and discussion}
\subsection{Growth and morphology}
The MnAs growth process is carried out by an MBE technique with use of an As$_2$ cracker source. The GaN substrate temperature is 350$^0$C. The Mn flux is calibrated by measuring RHEED oscillations for GaMnAs(100) calibration samples.\cite{sadowski00jvst} During the growth of the MnAs samples the Mn flux correspond to the layer growth rate of 0.8 ML/min for MnAs with the NiAs-type structure. As a consequence, in this paper we measure layer thickness in monolayers of relaxed hexagonal MnAs.

The growth of MnAs is monitored {\em in situ} by RHEED. For the clean GaN(000$\overline 1$)-(1x1) surface, a streaked pattern, characteristic of an atomically flat surface, is observed. The first layer of MnAs is grown by two different modes: with both Mn and As sources open (regular MBE growth) (I) or by an atomic layer epitaxy (ALE) - like technique (first an Mn layer is deposited then, the Mn source is shut and the As source is open) (II). Independently of the growth method, the first stages of deposition ( about 1 ML) cause blurring of the RHEED pattern. Further deposition, performed in a standard MBE mode, leads to improvement of the pattern (streaks became again stronger and sharper) - then, at about 5 ML, the critical thickness is achieved and the pattern switched to a dotted one, indicative of 3D growth. The mechanism of 2D growth mode observed, in spite of the large lattice mismatch in the system, for the thin MnAs layers on GaN seems to be similar to that described for InAs/GaAs.\cite{asklund01} The large lattice mismatch is partly relaxed by dislocations at the beginning of the growth process and this leads to the blurring of RHEED pattern observed for very thin layers. Further growth of MnAs, on top of the defected layers, results in formation of a smoother surface and the improved streaky RHEED pattern. When the epilayer becomes thicker than 5 ML, dots form on the surface.

\begin{figure}
\includegraphics[width=15cm]{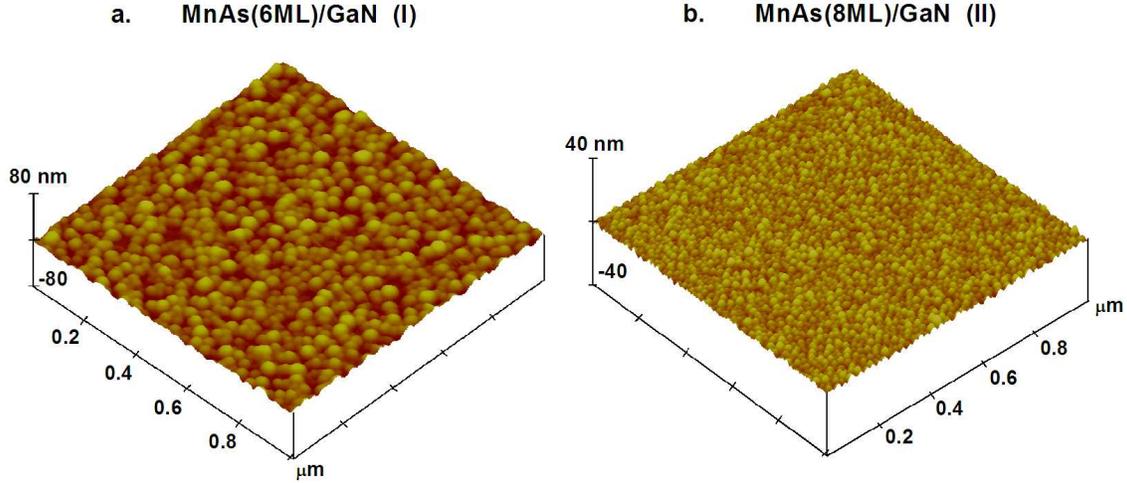}
\caption{\label{afm}The AFM images of MnAs(6 ML)/GaN(000$\overline
1$) (I) and MnAs(8 ML)/GaN(000$\overline 1$) (II) systems.}
\end{figure}

An AFM investigations of the samples morphology confirm that MnAs dots appear (Fig.~\ref{afm}) for both growth procedures. The sample prepared by the regular MBE deposition (with MnAs amount equivalent to 6 ML) is denoted by I. The obtained dots have the average diameter of 40 nm. After the ALE-like initiation of the growth followed by standard MBE deposition (the sample with 8 ML of MnAs -- II), the dots are somewhat smaller -- they have the average diameter of 25 nm. In both cases, the average height of the dots is equal to 4 nm. However, it was proved by Sadowski {\em at al.} \cite{sadowski05apl} that AFM markedly overestimates the size of dots in comparison with HRTEM results. The diameter of the dots determined by AFM did not correspond to their size but was a convolution of the real dot shape and the shape and size of the AFM tip. The dot diameter derived from HRTEM in Ref.~\onlinecite{sadowski05apl} was about 10 nm while that estimated from the AFM data was approximately eight times bigger. Therefore, the average density of dots seems to characterize better the dot systems and it allows reliable comparison of the results obtained by different groups. For sample I (Fig.~\ref{afm}a), the dot density is about 1$\cdot 10^{11}$ cm$^{-2}$, for sample II (Fig.~\ref{afm}b) -- $2.5\cdot 10^{11}$ cm$^{-2}$. For comparison, MnAs dots (with zinc blende or NiAs-type structure) on GaAs\cite{okabayashi04} formed with the density of 1.5--3.5$\cdot 10^{12}$ cm$^{-2}$, NiAs-type MnAs dots on annealed GaMnAs\cite{sadowski05apl} -- $2\cdot 10^{10}$ cm$^{-2}$, zinc blende Mn(Ga)As nanoclusters in GaAs\cite{moreno02} -- $1\cdot 10^{10}$ cm$^{-2}$. Clearly, the density of dots shown in Fig.~\ref{afm} corresponds well to the range of values reported in literature. Thus, we expect that the real average diameter of the dots does not deviate much from 10 nm, reported for MnAs dots by Sadowski {\em at al.} \cite{sadowski05apl} and Okabayashi {\em at al.} \cite{okabayashi04}.

\subsection{Resonant photoemission study}
Modification of the starting stage of MnAs growth does not result in qualitative changes in the morphology of the MnAs/GaN system. However, the corresponding electronic structures revealed by photoemission spectroscopy turn out to be markedly different. These studies give us a deeper insight into the structure properties of the MnAs part of the system. Since the contribution of Mn 3d states to the electronic structure of the system is the key issue which must be discussed, resonant photoemission technique has been applied. This method enables us to resolve contribution of 3d states of transition metal atoms to the photoemission spectra against the background emission from the valence band. The resonant photoemission experiments should be carried out for photon energies close to Mn~$3p \rightarrow 3d$ transition. When the photon energy matches the energy of this intra-ion transition to the unoccupied states in the partly filled shell (like Mn 3d), two processes leading to different final states of the same energy may occur.

\begin{figure}
\includegraphics[width=14cm]{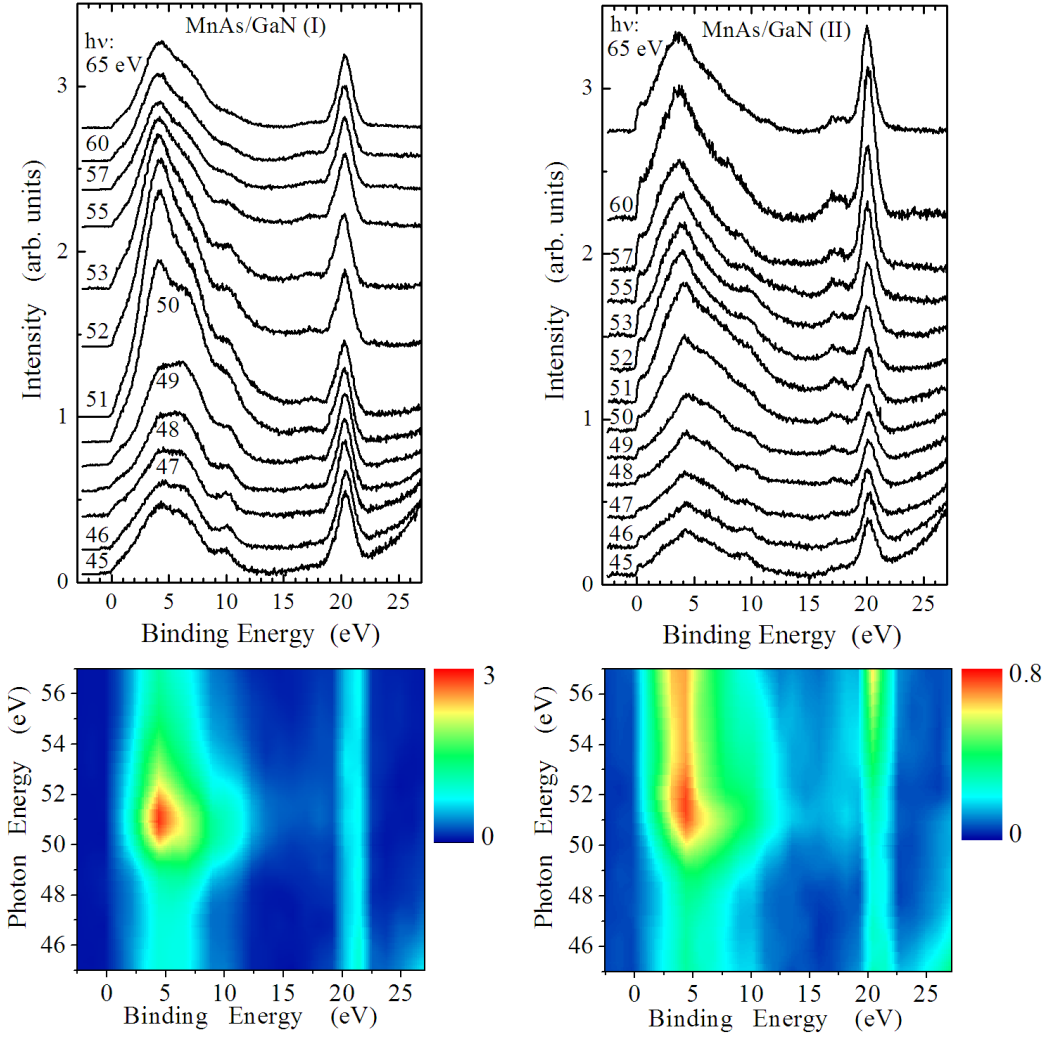}
\caption{\label{edc}The sets of energy distribution curves obtained for type I MnAs(6 ML)/GaN(000$\overline 1$) (a) and type II MnAs(8 ML)/GaN(000$\overline 1$) (b) for photon energies close to Mn~$3p \rightarrow 3d$ resonance.}
\end{figure}

The first one is the direct photoemission from the open shell to the continuum of free electron states. The second one is a process of the discrete intra-ion excitation. For Mn-containing crystals one considers the following transitions:

\begin{equation}\label{1}
    Mn 3d^{5} + h\nu \rightarrow Mn 3d^{4} + e^{-}
\end{equation}
    and
\begin{equation}\label{2}
       Mn 3p^{6}3d^{5} + h\nu \rightarrow Mn 3p^{5}3d^{6}.
\end{equation}
The Mn $3p^{5}3d^{6}$ state may autoionize, by super Coster-Kronig process, to Mn $3p^{6}3d^{4} + e^{-}$.
This leads to a Fano resonance \cite{fano61} and a Fano-type line appears in the absorption spectra at the energy corresponding to the Mn~$3p \rightarrow 3d$ transition. As a consequence, photoemission from the open shell is resonantly increased. Therefore the resonant photoemission is widely used for identification of the features in the photoemission spectra which can be ascribed to the emission from partly filled shells. In particular, a comparison of the spectra taken for the photon energies corresponding to the maximum (close to the resonance energy) and minimum (the anti-resonance energy at which the emission from the open shell is suppressed) of the Fano profile clearly shows enhancement of 3d--shell--related spectral features.

Fig.~\ref{edc} shows sets of photoemission spectra acquired at photon energies from 45 to 65 eV (i.e. covering the photon energy of the Mn~$3p \rightarrow 3d$ resonance -- about 51 eV). They have been measured for both types of  MnAs layers overgrown on GaN(000$\overline 1$)-(1x1). The binding energy range of the spectra, measured with respect to the Fermi energy, corresponds to the valence band (0-12~eV) and the Ga~3d maximum (at about 20~eV). Analysis of these sets of spectra shows that the intensity of emission from the valence band is strongly sensitive to photon energy. The resonance conditions occur for about 51 eV. The spectra taken at 48~eV correspond to the anti-resonance minimum of the Fano profile. Apart from these strong resonances at the same photon energies, both sets of spectra show important differences. The most important one occurs at the Fermi energy. For the sample II, the sharp edge is visible. For the sample I, the leading edge of the spectrum also starts at the Fermi energy but it is much less steep. This shows that the sample II has metallic properties, in contrast to the sample I. The overall shapes of the valence band main maxima (at about 4 eV) are also somewhat different and additional shoulders at 6 and 10~eV are clearly discernible for the sample I.

Let us turn to the maximum occurring at 20~eV. It is related to the Ga~3d core level. Its presence indicates that considerable emission from the GaN substrate contributes to the recorded spectra. It is also possible to detect differences between interfaces formed in both structures. In particular, the additional feature appearing for the sample II at about 17~eV, just above the Ga~3d peak, suggests different interactions occurring at the interface formed between GaN and MnAs.

\begin{figure}[h!]
\includegraphics[width=13cm]{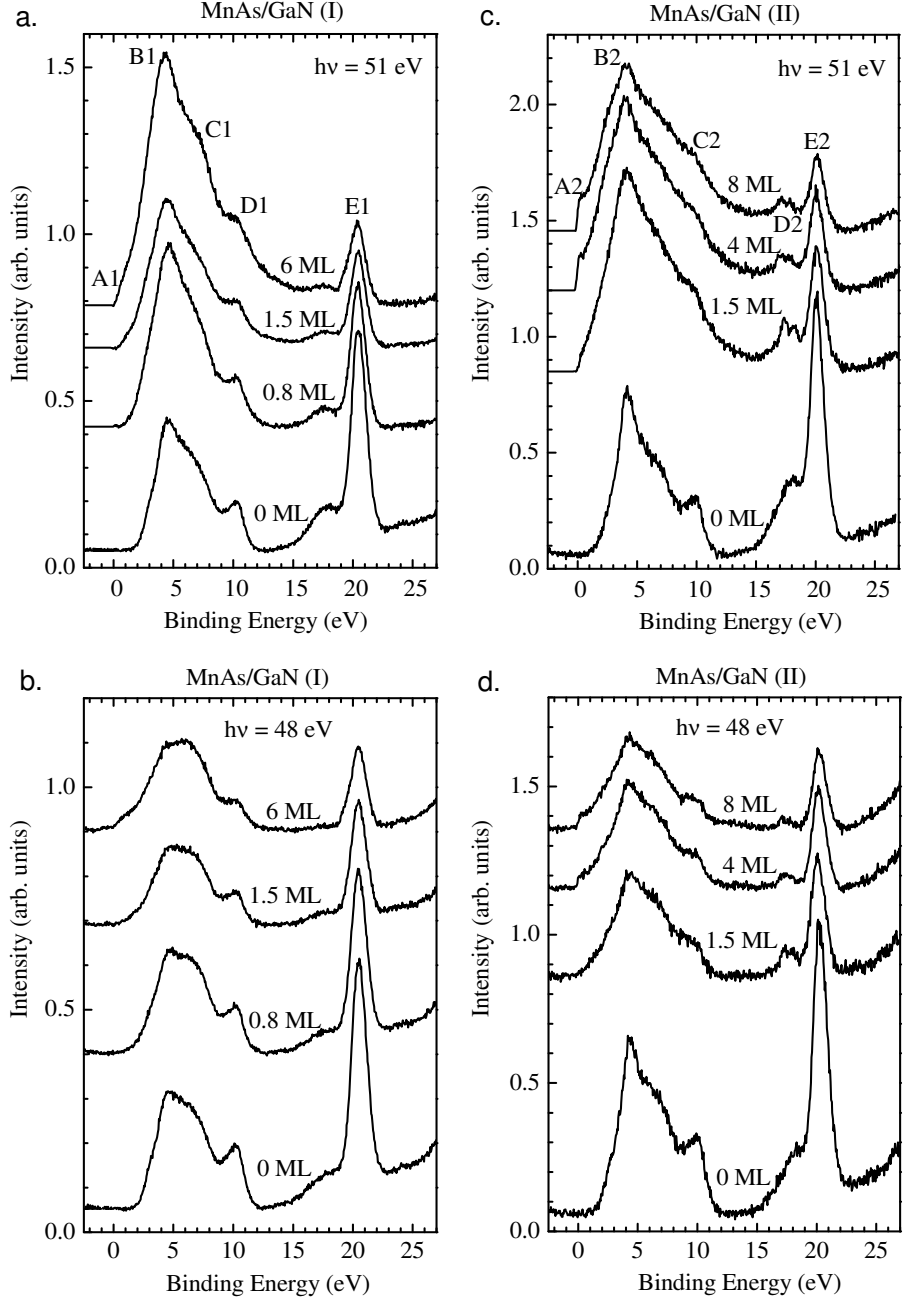}
\caption{\label{resares}Energy distribution curves taken for h$\nu$=51 eV (resonance) (a, c) and 48~eV (anti-resonance) (b, d)  after subsequent growth stages of type I MnAs/GaN(000$\overline 1$) (a, b) and type II MnAs/GaN(000$\overline 1$) (c, d) structures. MnAs coverage thickness is indicated at the graphs.}
\end{figure}

In order to visualize the development of the electronic structure of MnAs/GaN system and variations occurring due to different formation of the first monolayer of MnAs we show sets of photoemission spectra acquired for the samples I and II, under resonance and anti-resonance conditions, for clean GaN surface and for subsequent stages of MnAs layer growth (Fig.~\ref{resares}). The spectra obtained for clean surfaces match those previously reported for GaN crystal surfaces prepared by the same method, \cite{kowalski04ss} including small variations in relative intensities of the main maximum (at about 4 eV) and the shoulder at 7~eV. According to our previous studies, it is clean GaN(000$\overline 1$) surface with hexagonal (1x1) symmetry, showing surface electronic structure features characteristic of regions with the relaxed clean GaN$(000\overline 1)$ surface structure as well as of those with $(000\overline 1)$:Ga configuration, covered with additional layer of Ga atoms bound at the "on top" positions above N atoms.\cite{kowalski04ss} Difference between the spectra taken under Mn~$3p \rightarrow 3d$ resonance ($h\nu$=51~eV) and anti-resonance ($h\nu$=48~eV) conditions for clean GaN is negligible in comparison with those observed for samples covered with MnAs, independently of the thickness.
For both cases, first two stages of MnAs growth correspond to layer thickness lower than the critical one (5 ML), the third stage -- to the layer transformed into dots. Marked changes in the shape of the valence band spectrum appear as soon as the first layers of MnAs have been deposited. They manifested themselves both in the resonance and anti-resonance spectra. However, the main MnAs-related maximum, clearly visible in the resonance spectra, gets markedly decreased for the anti-resonance case. Intensity of MnAs-related features increase with rising MnAs layer thickness, in particular for the sample I, however MnAs dots formation does not lead to qualitative changes in the spectra. The intensity of Ga~3d core line decrease with of the MnAs layer nominal thickness. However, we want to point out that the strongest change occurs after a growth of the very first MnAs layers (about 1~ML). Further growth of MnAs causes relatively weak decrease of the maximum. The Ga~3d peak is still clearly discernible for the thickest MnAs layers (6, 8~ML). This confirms that MnAs does not form a uniform layer with constant thickness.

The main differences between the sample I and II, including the Fermi edge and the additional structure at 17--18~eV, are discernible in all spectra II for both photon energies. Intensity of the Fermi edge of the sample II increases with MnAs coverage. The feature close to the Ga~3d peak (D2) decreases together with suppression of the main Ga~3d peak (E2) due to increasing thickness of MnAs layer. This structure closely resembles that appearing after deposition of Mn on the GaN$(000\overline 1)$-(1x1) surface \cite{kowalik04} -- an additional component of the Ga~$3d$ spectral feature, at about 2 eV above the main peak. This shows that first layer of Mn, deposited in the ALE-like process, interacts with GaN, the surface gets disrupted and a reactive, Ga-rich interface is formed. Then, next layers of MnAs are grown and the intensity of this feature decreases. This indicates that the excessive Ga responsible for this structure remains at the interface.  For the sample I, no sings of surface disruption can be discerned in the spectra. Manganese and arsenic delivered to the clean GaN surface forms MnAs layer instantly, without anterior reacting with GaN substrate.

\begin{figure}
\includegraphics[width=16cm]{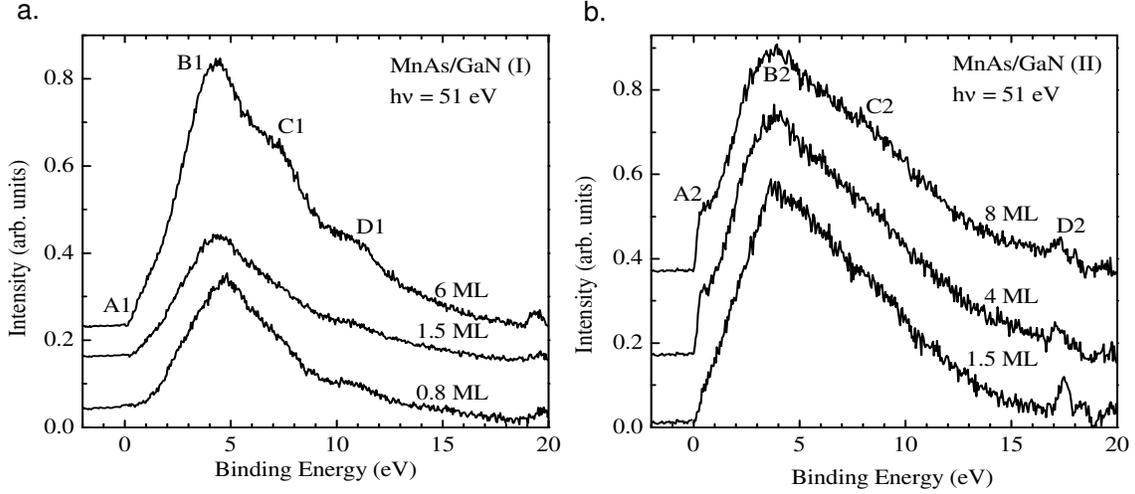}
\caption{\label{mnas}Emission from MnAs --- a measure of MnAs density of electronic states --- revealed by subtraction of the EDCs of clean GaN from the spectra of MnAs/GaN(000$\overline 1$) I (a) and II (b) after subsequent stages of the system formation. Curves are shifted along the intensity axis for clarity.}
\end{figure}

The considerable intensity of Ga~3d peak indicates that a part of the recorded spectra originates from the GaN substrate. In order to reveal emission from the MnAs dots we subtract the spectra taken for the respective GaN substrate. Before subtraction, the spectra of GaN have been normalized in an accordance with the intensities of the Ga~3d peaks. This enables us to take into account dumping of GaN emission increasing with MnAs overgrowth. Figs.~\ref{mnas} a and b show results of such procedure applied to the spectra taken at $h\nu~$=51~eV (the resonance conditions) for the samples I and II, respectively. They correspond mainly to emission from the valence band of MnAs but some features related to modifications of the substrate surface (like the feature D2 (17.2 and 18~eV for 1.5~ML MnAs) in Fig.~\ref{mnas}b) also occur. We assign them to the presence of excessive Ga at the interface due to interaction of Mn with GaN substrate.

The subtraction of the GaN substrate emission emphasizes the differences between both types of the samples indicating that MnAs systems formed on the samples I and II have markedly different character. Firstly, a sharp edge at the Fermi energy (A2), occurs only for the sample II. Secondly, the main feature of the valence band is sharper for the sample I and additional shoulders can be resolved on its high binding energy slope (C1 at 7.2 and D1 at 11~eV).

We find that the overall shape of the MnAs related emission from the sample II (Fig.~\ref{mnas}b) corresponds well to that of hexagonal, ferromagnetic $\alpha$-MnAs. Results of calculations \cite{shirai98} suggested that for this phase the main maximum of the spectrum occured at about 2~eV below the Fermi edge. However, photoemission experiments carried out for epilayers of hexagonal MnAs clearly showed that the main maximum of the spectrum taken for photon energy close to Mn~$3p \rightarrow 3d$ resonance should occur at about 4~eV below the clearly discernible, sharp Fermi edge \cite{shimada98,shimada99,okabayashi04,sadowski05apl}. This shape of the spectrum is indicative for metallic character of the system with relatively high density of states at the Fermi energy.

The shape of the spectra of the sample I seems to correspond to those reported for systems of dots of zinc-blende MnAs \cite{ono02,okabayashi04}. In particular, similarity of the energy position of the main maximum, very low emission intensity at the Fermi energy and additional shoulders appearing in the spectra obtained by Ono {\em et al.} \cite{ono02}, Okabayashi {\em et al.} \cite{okabayashi04} and in this work justify this observation.

However, according to the literature, the zinc-blende structure of MnAs dots, \cite{ono02,okabayashi04} or even layers, \cite{kim06} were stabilized by zinc-blende GaAs substrate. As Moreno {\em at al.} \cite{moreno02} and Okabayashi {\em et al.} \cite{okabayashi04} showed, zinc blende structure was sustained only in very small MnAs clusters or dots, with diameter of less that 10 nm (as measured by high-resolution transmission electron microscopy (HRTEM)). Bigger MnAs dots had NiAs-type structure.\cite{okabayashi04,moreno02,sadowski05apl} Since the GaN substrate has hexagonal structure, this mechanism cannot lead to the formation of zinc blende MnAs dots on the sample I. Besides, the dots on the sample I are bigger than those on the sample II (with NiAs-type structure).

The photoelectron spectroscopy probes directly the electronic states distribution in the system but not its crystallographic structure. Therefore, the results reported in this paper cannot prove whether the dots have zinc blende structure or NiAs-type one. Unfortunately, common X-ray diffraction methods does not give conclusive results for so thin MnAs structures \cite{paszkowicz05}. However, it was shown by theoretical calculations that half-metallic electronic structure with the Fermi level lying in the energy band gap of the minority-spin bands is likely to occur in zinc-blende MnAs \cite{shirai98jmm,zhao02,kim04,hong05} as well as in other manganese pnictides \cite{zheng04,hong05}. Distortions of the crystal structure may also influence band structure of these materials and their magnetic properties. Tetragonal distortions of zinc-blende structure in MnSb or MnBi support half-metallicity and ferromagnetism \cite{zheng04} or, in MnAs/Si, such distortions induce transition to ferromagnetism and half-metallicity \cite{kim04}. In contrast to that, for hexagonal MnAs, Debernardi {\em et al.} \cite{debernardi03cms,debernardi03mse} showed that uniaxial stress in NiAs-type MnAs (related to pseudomorfic growth on a lattice mismatched substrate) led to some non-linear changes in electronic and magnetic properties, but they did not result in half-metallic electronic band structure. Therefore, formation of MnAs with zinc blende crystallographic structure still seems to be a necessary condition for occurrence of half-metallic band structure in the system. However, such an interpretation of our results needs confirmation that a cubic phase of the hexagonal crystal can grow on the hexagonal substrate.

We believe that tentative arguments for it can be derived from a phenomenon reported by Kitamura {\em et al.} \cite{kitamura06} for InGaN epilayers grown on InN. For In contents below 0.66, a cubic phase appeared in the InGaN epilayers thicker than a critical thickness. That was ascribed to degradation of the crystalline quality due to an abrupt change in the crystal composition at the interface. MnAs/GaN would be another system in which such phenomenon could occur. We postulate that differences in initial growth stages of the sample I and II and different interactions between Mn and the GaN substrate lead to different crystalline structures of the overgrown MnAs. The interface of the sample I seems to be more abrupt due to the absence of interaction between components of MnAs and the GaN substrate. If this promotes cubic phase of MnAs formation, half-metallic electronic structure may occur and manifest itself in the reported photoemission experiments. Confirmation of this suggestion needs further studies, in particular crystal structure investigations by means of surface sensitive methods.

\begin{figure}
\includegraphics[width=10cm]{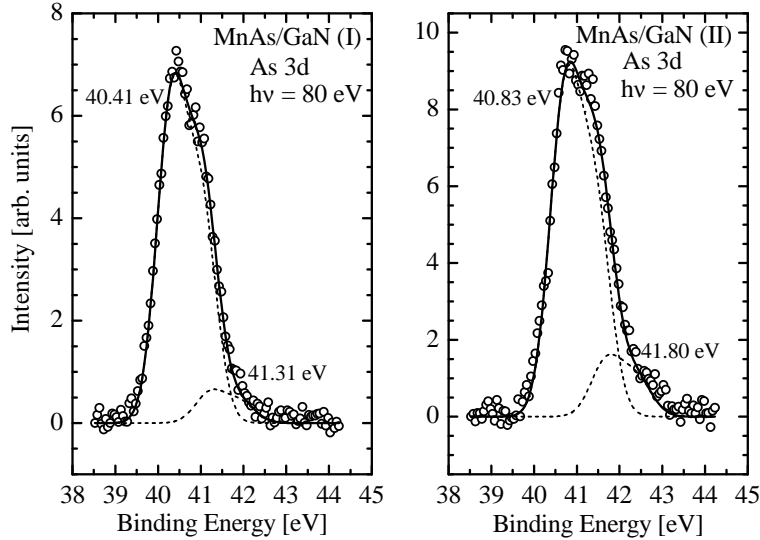}
\caption{\label{as3d}The As~3d feature measured for MnAs(6 ML)/GaN(000$\overline 1$) (I) and MnAs(8 ML)/GaN(000$\overline 1$) (II) shown together with the results of deconvolution into spin-orbit split dublets of  Gaussians.}
\end{figure}

The next argument for the difference between structures of MnAs(I) and MnAs(II) arises from the photoelectron spectroscopy of the As 3d core level. In order to reveal the energy shifts of this core level, reflecting changes in chemical bonding and structure of MnAs, As~3d maxima have been recorded at $h\nu~$=80~eV for both systems (Fig.~\ref{as3d}). For the energy scale with origin at the Fermi energy, the Ga~3d peaks of the sample I and II were aligned well but for the As~3d feature an energy shift of about 0.4~eV clearly occurred. The same energy position of the Ga~3d peak observed for both samples proved that surface conditions on them are similar. The energy shift of As~3d feature can be interpreted as a manifestation of different structure of the MnAs layers, consistent with differences revealed in the MnAs valence band shape.

Each of the As 3d features can be decomposed into two Gaussian dublets corresponding to As 3d$_{5/2}$ and As 3d$_{3/2}$ with spin-orbit splitting of 0.7 eV \cite{eastman80,prietsch88}. The dublets are separated by about 0.9~eV. The second one, at the higher binding energy, is markedly weaker and corresponds to 10-15 \% of the total intensity of the As~3d feature, for both systems. Similar energy separation and relative intensity occured for thinner MnAs layers. Therefore, we cannot interpret the minor component as corresponding to surface As atoms. Moreover, we could expect that surface related contribution to the As~3d maximum would be shifted to lower binding energies with respect to the bulk related one, like in GaAs \cite{eastman80,prietsch88,chaika99}. Similarly, this feature cannot correspond to As atoms in a different lattice position or chemical state. It is hardly possible that relative populations of As atoms in different states are independent from the MnAs layer thickness, structure and morphology. Thus, we can only interpret it tentatively as a satellite-like structure.

\begin{figure}
\includegraphics[width=10cm]{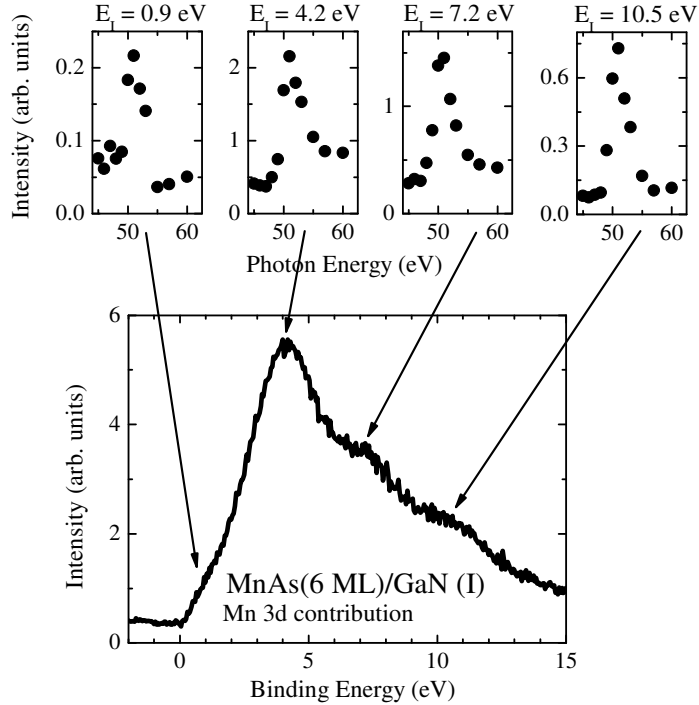}
\caption{\label{cis8}Difference spectrum --- a measure of Mn~3d contribution to the valence band --- obtained by subtraction of the anti-resonance spectra ($h\nu$=48~eV) from the resonance ($h\nu$=51~eV) ones for type I MnAs/GaN(000$\overline 1$). Constant initial state spectra were made for the initial state binding energies corresponding to the main features of the difference spectrum.}
\end{figure}

\begin{figure}
\includegraphics[width=9cm]{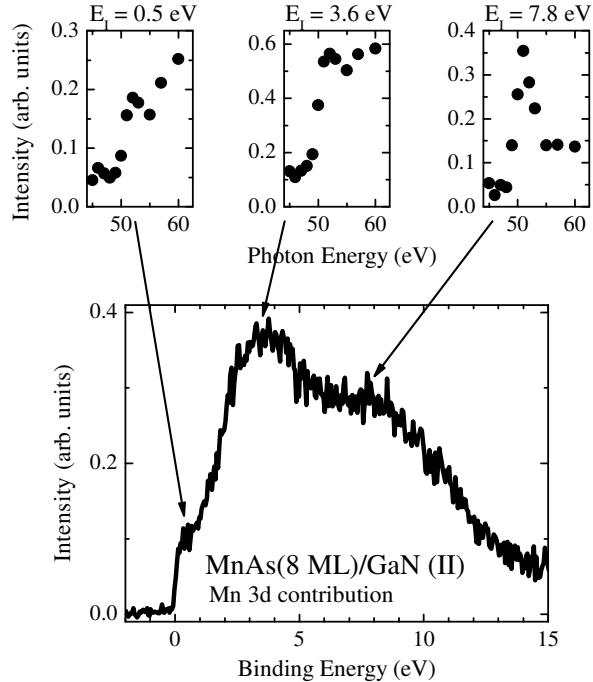}
\caption{\label{cis10}Difference spectrum obtained by subtraction of the anti-resonance spectra ($h\nu$=48~eV) from the resonance ($h\nu$=51~eV) ones for type II MnAs/GaN(000$\overline 1$). Constant initial state spectra were made for the initial state binding energies corresponding to the main features of the difference spectrum.}
\end{figure}

Further information confirming the difference between electronic structures of the two investigated MnAs systems has been obtained thanks to one of the most important advantages of resonant photoemission spectroscopy --- its enhanced sensitivity to emission from open shells of transition elements. Figs.~\ref{cis8} and \ref{cis10} show difference spectra obtained by subtraction of curves obtained for anti-resonance photon energy from those taken under resonance conditions. In such spectra which measure Mn~3d contribution to the valence band, the main differences between the sample I and II manifested themselves even more discernibly. Consistently with the results of the data analysis presented above, the curve obtained for the sample I is very similar to the difference spectra of zinc-blende MnAs dots, reported by Ono {\em et al.} \cite{ono02}. Only the shoulder at 7.2~eV is more pronounced in our spectrum.

The differences between MnAs I and II manifest themselves explicitly in the constant-initial-state (CIS) photoemission spectra (Figs.~\ref{cis8} and \ref{cis10}). They show photoemission intensity (relevant GaN substrate emission subtracted) as a function of photon energy for the initial state binding energies corresponding to main features of the difference (resonance -- anti-resonance) spectra. For the sample I (Fig.~\ref{cis8}), the curves obtained for four features at 0.9, 4.2, 7.2 and 10.5~eV differ in intensity but a sharp maximum at the same photon energy can easily be seen in each of them. For the sample II (Fig.~\ref{cis10}), a sharp peak occurs only in the CIS curve corresponding to the shoulder at 7.8~eV. For the main maximum and the shoulder just below the Fermi edge, only very weak features on a non-resonant background appear. This set of CIS spectra corresponds well to those reported for the systems exhibiting in photoemission a strongly resonant satellite and an almost non-resonant main line, like Ni \cite{barth79}. Such character of photoemission spectrum is interpreted as a manifestation of two different final state occurring in the system. The main line corresponds to a final state fully screened in the d-shell, while the satellite --- to that screened by band electrons (4s in nickel). However, the CIS spectra of the sample I can be compared with those of many Mn compounds, like MnCl$_{2}$ \cite{kakizaki83} or diluted magnetic semiconductors with Mn \cite{ley87}. For those systems, Fano-like lines with distinct maxima were observed in CIS spectra throughout the d-shell related difference spectrum. Analysis of the results collected for various transition-metal-containing compounds showed that appearance of a resonant satellite with a non-resonant main line in the spectra corresponded to large configuration interaction or strong mixing of d and ligand states in the system \cite{davis86}. Thus, we can expect that similar difference in these characteristics occurs also in the samples I and II. This supports our supposition that the surroundings of Mn atoms in the investigated systems differ substantially.

\begin{figure}
\includegraphics[width=15cm]{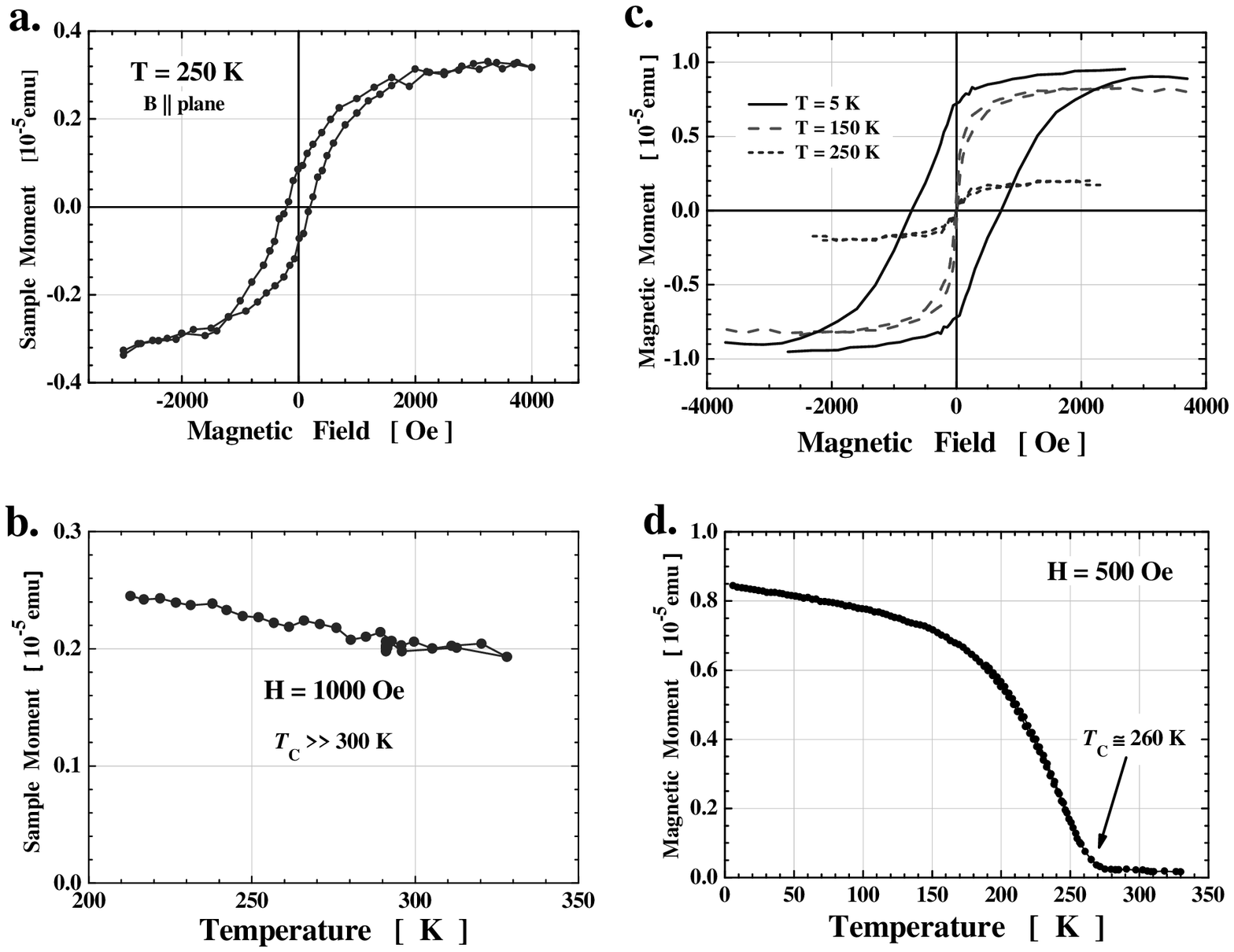}
\caption{\label{magn}Magnetic moment of type I (a,b) and type II (c,d) MnAs nano-dots. Diamagnetic (linear in field) GaN substrate signal has been subtrated for the data.}
\end{figure}

\subsection{Magnetic properties}
Magnetic properties of the MnAs dots are investigated in a home made SQUID magnetometer.  Fig.~\ref{magn} shows the magnetic field, m(H), and the temperature dependence, m(T), of the net magnetic moment of both types of MnAs structures. All the measurements are performed with the magnetic field applied in the plane of the samples. Both systems exhibit strong nonlinear magnetic field response even at high temperatures and show a development of coercivity and remanence on lowering T. It is however important to stress it here that neither of these, nor even the clear saturation of the magnetization, can be taken as an indication of a presence of a uniform ferromagnetic state. On the contrary, the characteristic rounded shape of M(H) curves (Langevin function alike) suggests that we deal with predominantly superparamagnetic behavior of an ensemble of ferromagnetic particles, but the small magnetic remanence (or coercivity) indicates a presence of strong enough magnetic anisotropy to block (at least partially) some of the particles even at temperatures as high as 250 K. Therefore an external non-zero magnetic field is required to flip the moments of those nano-dots which got trapped behind the energy barrier for their magnetization reversal:  $\Delta E = KV$, where $V$ is the particle volume, and $K$ is the relevant magnetic anisotropy energy density. This leads to non-zero coercivity and remanence, resulting in an open magnetic hysteresis of the system. The maximum temperature at which on the time scale of the experiment, the statistically significant number of magnetic moments remains blocked is called the blocking temperature, T$_{B}$, however, a minute blocking effects can persist to much higher temperatures, T$_{B}^{max}$, if the particle size/anisotropy distribution is particularly large. We note that, a similar behavior can be brought about if there exists a substantial magnetic coupling between the particles. However, observed here varying width of the m(H) curves, with the most narrow point at the magnetization reversal (seen particularly in Fig.~\ref{magn}c) indicates strongly the dominance of activated processes related to the existence of the barriers over the mutual magnetic coupling in the investigated nano-dots systems.

It is widely accepted that for a system of identical non-interacting particles the following relationship defines T$_{B}$:
\begin{equation}\label{3}
    T_B \simeq KV/25k_{B}
\end{equation}
and so for real systems a maximum temperature in which blocking effects are observed serves as an upper limit for $KV$ product. Alternatively, knowing $K$ or $V$ beforehand, the relationship allows to evaluate the other. As argued in the section A, an average MaAs dot diameter is expected to be ~10 nm. We can independently confirm this finding by fitting the Langevin function to the 250 K data for type II hexagonal MnAs dots, for which a near perfect superparamagnetic behavior is seen. An excellent fit (not shown for clarity on Fig.~\ref{magn}c) is obtained assuming an effective magnetic moment  $\mu = 40 000 \mu_{B}$ for each MnAs dot. Since at 250 K MnAs saturation magnetization is approximately 2 $\mu_{B}$ per Mn atom, the effective "magnetic" volume of an average MnAs nano-dot approximates to 1000 nm$^{3}$, giving the lateral size of 10 nm, as deduced from AFM studies. Now employing Eq.~\ref{3} we arrive at $K \simeq 1 \cdot 10^{5}$ erg/cm$^3$ required to reproduce T$_{B}$ of 200 K as results in Fig.~\ref{magn}c suggest. This is a rather large value for $K$ laying between anisotropy constants of iron (cubic) and cobalt (uniaxial), and is approximately 20 times that of bulk MnAs (uniaxial) \cite{ploog06}. But such a discrepancy can by no means devalue the proposed analysis. As argued above, this is the upper limit for $K$, with no solid indication how many times the target is overshoot. It also important to bear in mind that the nano-dots we investigate are formed only owing to a large epitaxial/misfit strain, and its presence is expected to augment the magnetic anisotropy substantially.

A similar analysis for type I nano-dots is largely precluded since we do not posses any m(H) dependency for this system in the pure superparamagnetic regime, see Fig.~\ref{magn}a. It is likely that for these nano-dots the blocking is comparable with, or even greater than their intra-dot Curie temperature, T$_{C}$, approximated later to a value rather closer to 400 K than to typical 313 K for free standing bulk MnAs. One of the reasons for such an enlargement of T$_{B}$ can be a larger volume of these dots. We recall here that the AFM morphology studies suggest their 2.5 smaller surface density what can be translated into 4 times greater average volume than for the type II nano-dots. Now, taking similarly enlarged magnetic anisotropy, T$_{B}$ values of even a few hundred K can be easily accounted for. The considered in Ref.~\onlinecite{kowalski04app} contribution of hypothetic manganese compounds with a very high T$_{C}$, formed at the interface, can be excluded because they would manifest themselves even stronger in the sample II, with interaction of Mn with GaN clearly revealed by spectroscopy of Ga~3d core level.

Because the structures behave superparamagnetically, neither the upper temperature limit for non-zero coercive force (T$_{B}^{max}$) nor the vanishing of the remnant magnetization is a appropriate measure of the intra-dot T$_{C}$, which on the account of the data presented in Fig.~\ref{magn}a,c obviously must be higher than any of these estimates. To determine T$_{C}$ we investigate m(T) dependency at nearly saturating magnetic field, see Fig.~\ref{magn}b,d. The two different field values used in these experiments are compromised to satisfy both the requirement to be significantly close to saturation and to minimize increasing with the applied field noise in the magnetometer. Under these experimental conditions the magnetization for type I MnAs dots remains clearly non-zero up to the highest available temperature in out magnetometer. Additionally, its very weak temperature dependence indicates rather high, substantially above 330 K, intra-dot T$_{C}$, so exceeding considerably that of the bulk MnAs (313 K). Markedly different behavior is seen for the second structure. Here, in the whole temperature range, the nearly saturated magnetization follows reasonable well the Brillouin function, yielding the intra-dot T$_{C}$ only of about 260 K. In turn, this value is considerably lower than that for the bulk MnAs. Although we can not propose a convincing microscopic arguments accounting for such a huge, at least 100 K difference in T$_{C}$, we note that similarly enlarged intra-dot T$_{C}$ values for zinc-blende MnAs nano-dots can be inferred from Ono {\em et al.} \cite{ono02} and Moreno {\em et al.} \cite{moreno02} magnetic data \footnote{In Refs.~\onlinecite{moreno02} and \onlinecite{ono02} Curie temperature was assessed from the upper temperature limit from either non-zero remnant magnetization or non-zero coercivity, respectively. As argued in this paragraph this is at the best only the lower limit for the intra-particle (dot) T$_{C}$, particularly if the particles are randomly oriented and strained by the host environment. Indeed if our arguments are applied, the abovementioned experiments yield much higher values of T$_{C}$.}, despite different substrate/host matrix in their studies. So, this seems to be a common feature for zinc-blende MnAs dots. For the hexagonal, type II MnAs nano-dots, a T$_{C}$ lowering seems to be connected with layer-to-(nano-)dots transition and a smaller size of these dots. Most likely, an increased number of surface moments in comparison to the inside ones in the smaller dots reduces the overall strength of the magnetic coupling, resulting in noticeable T$_{C}$ lowering.

\section{Conclusions}
MnAs dots have been grown on the GaN(000$\overline 1$)-(1x1) surface by means of MBE. The dots form spontaneously in the layers thicker than 5 ML. Two methods of growth initiation have been tested: regular MBE and ALE-like (with noticeable reaction between manganese and GaN substrate). The structures grown by these methods have similar morphology but markedly different electronic structures. The ALE-like procedure leads to the electronic structure characteristic of ferromagnetic NiAs-type MnAs. The regular MBE growth of the layer results in the density of states distribution suggesting half-metallic character of the MnAs dots. Magnetization measurements revealed further differences between the systems under investigation. Both of them exhibit superparamagnetic properties but the basic parameters describing them are considerably different. For MnAs(II), the blocking temperature, deduced from the m(H) curves, is about 200 K, the intra-dot T$_{C}$ --- about 260 K. For MnAs(I), such temperatures are markedly higher than 330 K. The upper-limit value for the anisotropy energy density ($K$) required to reproduce experimental results for MnAs(II) turned out to be unexpectedly much higher than that characteristic of bulk MnAs, probably due to the presence of large strain in the investigated dots.

Modification of the initial stage of MnAs layer growth on GaN turns out to be an effective method for changing electronic and magnetic properties of the system. Our results suggest that MnAs/GaN is a system for which an attempt to obtain a combination of half-metallic electronic structure and T$_{C}$ higher that room temperature could be reasonable. In such a material, spin-polarized currents can be effectively induced and it would be particularly useful for fabrication of spintronic devices.

\begin{acknowledgments}
This work was supported by MSHE (Poland) projects 1P03B 116 28 and N202 101 31/0749, by the European Community via "ASPECT" Centre of Excellence (Contract G6MA-CT2002-0421) and the Research Infrastructure Action under the FP6 "Structuring the European Research Area" Programme (through the Integrated Infrastructure Initiative "Integrating Activity on Synchrotron and Free Electron Laser Science") at MAX-lab.
\end{acknowledgments}

\bibliography{mnas}

\begin{thebibliography}{58}
\expandafter\ifx\csname natexlab\endcsname\relax\def\natexlab#1{#1}\fi
\expandafter\ifx\csname bibnamefont\endcsname\relax
  \def\bibnamefont#1{#1}\fi
\expandafter\ifx\csname bibfnamefont\endcsname\relax
  \def\bibfnamefont#1{#1}\fi
\expandafter\ifx\csname citenamefont\endcsname\relax
  \def\citenamefont#1{#1}\fi
\expandafter\ifx\csname url\endcsname\relax
  \def\url#1{\texttt{#1}}\fi
\expandafter\ifx\csname urlprefix\endcsname\relax\def\urlprefix{URL }\fi
\providecommand{\bibinfo}[2]{#2}
\providecommand{\eprint}[2][]{\url{#2}}

\bibitem[{\citenamefont{Heusler}(1904)}]{heusler04}
\bibinfo{author}{\bibfnamefont{F.}~\bibnamefont{Heusler}},
  \bibinfo{journal}{Z.\ Angew.\ Chem.} \textbf{\bibinfo{volume}{17}},
  \bibinfo{pages}{260} (\bibinfo{year}{1904}).

\bibitem[{\citenamefont{Blois and Rodbell}(1963)}]{deblois63}
\bibinfo{author}{\bibfnamefont{R.~D.} \bibnamefont{Blois}} \bibnamefont{and}
  \bibinfo{author}{\bibfnamefont{D.}~\bibnamefont{Rodbell}},
  \bibinfo{journal}{Phys.\ Rev.} \textbf{\bibinfo{volume}{130}},
  \bibinfo{pages}{1347} (\bibinfo{year}{1963}).

\bibitem[{\citenamefont{Wills and Rooksby}(1954)}]{wills54}
\bibinfo{author}{\bibfnamefont{B.}~\bibnamefont{Wills}} \bibnamefont{and}
  \bibinfo{author}{\bibfnamefont{H.}~\bibnamefont{Rooksby}},
  \bibinfo{journal}{Proc.\ Phys.\ Soc.\ London, Sect.\ B}
  \textbf{\bibinfo{volume}{67}}, \bibinfo{pages}{290} (\bibinfo{year}{1954}).

\bibitem[{\citenamefont{Goodenough}(1963)}]{goodenough63}
\bibinfo{author}{\bibfnamefont{J.}~\bibnamefont{Goodenough}},
  \emph{\bibinfo{title}{Magnetism and the Chemical Bond}}
  (\bibinfo{publisher}{Interscience, New York}, \bibinfo{year}{1963}).

\bibitem[{\citenamefont{Albers and Haas}(1964)}]{albers64}
\bibinfo{author}{\bibfnamefont{W.}~\bibnamefont{Albers}} \bibnamefont{and}
  \bibinfo{author}{\bibfnamefont{C.}~\bibnamefont{Haas}},
  \bibinfo{journal}{Phys.\ Lett.} \textbf{\bibinfo{volume}{8}},
  \bibinfo{pages}{300} (\bibinfo{year}{1964}).

\bibitem[{\citenamefont{B{\"a}rner}(1977)}]{barner77}
\bibinfo{author}{\bibfnamefont{K.}~\bibnamefont{B{\"a}rner}},
  \bibinfo{journal}{Phys.\ Stat.\ Solidi B} \textbf{\bibinfo{volume}{84}},
  \bibinfo{pages}{385} (\bibinfo{year}{1977}).

\bibitem[{\citenamefont{Sandratskii et~al.}(1981)\citenamefont{Sandratskii,
  Egorov, and Berdyshev}}]{sandratskii81}
\bibinfo{author}{\bibfnamefont{L.}~\bibnamefont{Sandratskii}},
  \bibinfo{author}{\bibfnamefont{R.}~\bibnamefont{Egorov}}, \bibnamefont{and}
  \bibinfo{author}{\bibfnamefont{A.}~\bibnamefont{Berdyshev}},
  \bibinfo{journal}{Phys.\ Stat.\ Solidi B} \textbf{\bibinfo{volume}{103}},
  \bibinfo{pages}{511} (\bibinfo{year}{1981}).

\bibitem[{\citenamefont{Podloucky}(1984{\natexlab{a}})}]{podloucky84jmm}
\bibinfo{author}{\bibfnamefont{R.}~\bibnamefont{Podloucky}},
  \bibinfo{journal}{J.\ Magn.\ Magn.\ Mat.} \textbf{\bibinfo{volume}{43}},
  \bibinfo{pages}{204} (\bibinfo{year}{1984}{\natexlab{a}}).

\bibitem[{\citenamefont{Podloucky}(1984{\natexlab{b}})}]{podloucky84ssc}
\bibinfo{author}{\bibfnamefont{R.}~\bibnamefont{Podloucky}},
  \bibinfo{journal}{Solid State Commun.} \textbf{\bibinfo{volume}{50}},
  \bibinfo{pages}{763} (\bibinfo{year}{1984}{\natexlab{b}}).

\bibitem[{\citenamefont{Coehoorn and de~Groot}(1985)}]{coehoorn85}
\bibinfo{author}{\bibfnamefont{R.}~\bibnamefont{Coehoorn}} \bibnamefont{and}
  \bibinfo{author}{\bibfnamefont{R.}~\bibnamefont{de~Groot}},
  \bibinfo{journal}{J.\ Phys.\ F: Met.\ Phys.} \textbf{\bibinfo{volume}{15}},
  \bibinfo{pages}{2135} (\bibinfo{year}{1985}).

\bibitem[{\citenamefont{Motizuki}(1987)}]{motizuki87}
\bibinfo{author}{\bibfnamefont{K.}~\bibnamefont{Motizuki}},
  \bibinfo{journal}{J.\ Magn.\ Magn.\ Mat.} \textbf{\bibinfo{volume}{70}},
  \bibinfo{pages}{1} (\bibinfo{year}{1987}).

\bibitem[{\citenamefont{Shirai and Tokioka}(1998)}]{shirai98}
\bibinfo{author}{\bibfnamefont{M.}~\bibnamefont{Shirai}} \bibnamefont{and}
  \bibinfo{author}{\bibfnamefont{Y.}~\bibnamefont{Tokioka}},
  \bibinfo{journal}{J. Electron Spectr.\ Rel.\ Phenom.}
  \textbf{\bibinfo{volume}{88-91}}, \bibinfo{pages}{357}
  (\bibinfo{year}{1998}).

\bibitem[{\citenamefont{Ravindran et~al.}(1999)\citenamefont{Ravindran, Delin,
  James, Johansson, Wills, Ahuja, and Eriksson}}]{ravindran99}
\bibinfo{author}{\bibfnamefont{P.}~\bibnamefont{Ravindran}},
  \bibinfo{author}{\bibfnamefont{A.}~\bibnamefont{Delin}},
  \bibinfo{author}{\bibfnamefont{P.}~\bibnamefont{James}},
  \bibinfo{author}{\bibfnamefont{B.}~\bibnamefont{Johansson}},
  \bibinfo{author}{\bibfnamefont{J.}~\bibnamefont{Wills}},
  \bibinfo{author}{\bibfnamefont{R.}~\bibnamefont{Ahuja}}, \bibnamefont{and}
  \bibinfo{author}{\bibfnamefont{O.}~\bibnamefont{Eriksson}},
  \bibinfo{journal}{Phys.\ Rev.\ B} \textbf{\bibinfo{volume}{59}},
  \bibinfo{pages}{15680} (\bibinfo{year}{1999}).

\bibitem[{\citenamefont{Bouwma and Hass}(1973)}]{bouwma73}
\bibinfo{author}{\bibfnamefont{J.}~\bibnamefont{Bouwma}} \bibnamefont{and}
  \bibinfo{author}{\bibfnamefont{C.}~\bibnamefont{Hass}},
  \bibinfo{journal}{Phys.\ Status Solidi B} \textbf{\bibinfo{volume}{56}},
  \bibinfo{pages}{299} (\bibinfo{year}{1973}).

\bibitem[{\citenamefont{Liang and Chen}(1977)}]{liang77}
\bibinfo{author}{\bibfnamefont{K.}~\bibnamefont{Liang}} \bibnamefont{and}
  \bibinfo{author}{\bibfnamefont{T.}~\bibnamefont{Chen}},
  \bibinfo{journal}{Solid State Commun.} \textbf{\bibinfo{volume}{23}},
  \bibinfo{pages}{975} (\bibinfo{year}{1977}).

\bibitem[{\citenamefont{Chen et~al.}(1976)\citenamefont{Chen, Stutius, Allen,
  and Stewart}}]{chen76}
\bibinfo{author}{\bibfnamefont{T.}~\bibnamefont{Chen}},
  \bibinfo{author}{\bibfnamefont{W.}~\bibnamefont{Stutius}},
  \bibinfo{author}{\bibfnamefont{J.}~\bibnamefont{Allen}}, \bibnamefont{and}
  \bibinfo{author}{\bibfnamefont{G.}~\bibnamefont{Stewart}}, in
  \emph{\bibinfo{booktitle}{Magnetism and Magnetic Materials}}, edited by
  \bibinfo{editor}{\bibfnamefont{J.}~\bibnamefont{Becker}},
  \bibinfo{editor}{\bibfnamefont{G.}~\bibnamefont{Lander}}, \bibnamefont{and}
  \bibinfo{editor}{\bibfnamefont{J.}~\bibnamefont{Ryne}},
  \bibinfo{organization}{AIP Conf.\ Proc.\ No.\ 29} (\bibinfo{publisher}{AIP},
  \bibinfo{address}{New York}, \bibinfo{year}{1976}), p. \bibinfo{pages}{532}.

\bibitem[{\citenamefont{Prinz}(1990)}]{prinz90}
\bibinfo{author}{\bibfnamefont{G.}~\bibnamefont{Prinz}},
  \bibinfo{journal}{Science} \textbf{\bibinfo{volume}{250}},
  \bibinfo{pages}{1092} (\bibinfo{year}{1990}).

\bibitem[{\citenamefont{Tanaka et~al.}(1993)\citenamefont{Tanaka, Harbison,
  Sands, Philips, Cheeks, Boeck, Florez, and Keramidas}}]{tanaka93}
\bibinfo{author}{\bibfnamefont{M.}~\bibnamefont{Tanaka}},
  \bibinfo{author}{\bibfnamefont{J.}~\bibnamefont{Harbison}},
  \bibinfo{author}{\bibfnamefont{T.}~\bibnamefont{Sands}},
  \bibinfo{author}{\bibfnamefont{B.}~\bibnamefont{Philips}},
  \bibinfo{author}{\bibfnamefont{T.}~\bibnamefont{Cheeks}},
  \bibinfo{author}{\bibfnamefont{J.~D.} \bibnamefont{Boeck}},
  \bibinfo{author}{\bibfnamefont{L.}~\bibnamefont{Florez}}, \bibnamefont{and}
  \bibinfo{author}{\bibfnamefont{V.}~\bibnamefont{Keramidas}},
  \bibinfo{journal}{Appl.\ Phys.\ Lett.} \textbf{\bibinfo{volume}{63}},
  \bibinfo{pages}{696} (\bibinfo{year}{1993}).

\bibitem[{\citenamefont{Kioseoglou et~al.}(2002)\citenamefont{Kioseoglou, Kim,
  Soo, Chen, Luo, Kao, Sasaki, Liu, and Furdyna}}]{kioseoglu02}
\bibinfo{author}{\bibfnamefont{G.}~\bibnamefont{Kioseoglou}},
  \bibinfo{author}{\bibfnamefont{S.}~\bibnamefont{Kim}},
  \bibinfo{author}{\bibfnamefont{Y.}~\bibnamefont{Soo}},
  \bibinfo{author}{\bibfnamefont{X.}~\bibnamefont{Chen}},
  \bibinfo{author}{\bibfnamefont{H.}~\bibnamefont{Luo}},
  \bibinfo{author}{\bibfnamefont{Y.}~\bibnamefont{Kao}},
  \bibinfo{author}{\bibfnamefont{Y.}~\bibnamefont{Sasaki}},
  \bibinfo{author}{\bibfnamefont{X.}~\bibnamefont{Liu}}, \bibnamefont{and}
  \bibinfo{author}{\bibfnamefont{J.}~\bibnamefont{Furdyna}},
  \bibinfo{journal}{Appl.\ Phys.\ Lett.} \textbf{\bibinfo{volume}{80}},
  \bibinfo{pages}{1150} (\bibinfo{year}{2002}).

\bibitem[{\citenamefont{Akeura et~al.}(1995)\citenamefont{Akeura, Tanaka, Ueki,
  and Nishinaga}}]{akeura95}
\bibinfo{author}{\bibfnamefont{K.}~\bibnamefont{Akeura}},
  \bibinfo{author}{\bibfnamefont{M.}~\bibnamefont{Tanaka}},
  \bibinfo{author}{\bibfnamefont{M.}~\bibnamefont{Ueki}}, \bibnamefont{and}
  \bibinfo{author}{\bibfnamefont{T.}~\bibnamefont{Nishinaga}},
  \bibinfo{journal}{Appl.\ Phys.\ Lett.} \textbf{\bibinfo{volume}{67}},
  \bibinfo{pages}{3349} (\bibinfo{year}{1995}).

\bibitem[{\citenamefont{Tanaka et~al.}(1999)\citenamefont{Tanaka, Saito, and
  Nishinaga}}]{tanaka99}
\bibinfo{author}{\bibfnamefont{M.}~\bibnamefont{Tanaka}},
  \bibinfo{author}{\bibfnamefont{K.}~\bibnamefont{Saito}}, \bibnamefont{and}
  \bibinfo{author}{\bibfnamefont{T.}~\bibnamefont{Nishinaga}},
  \bibinfo{journal}{Appl.\ Phys.\ Lett.} \textbf{\bibinfo{volume}{74}},
  \bibinfo{pages}{64} (\bibinfo{year}{1999}).

\bibitem[{\citenamefont{Ramsteiner et~al.}(2002)\citenamefont{Ramsteiner, Hao,
  Kawaharazuka, Zhu, K{\"a}stner, Hey, D{\"a}weritz, Grahn, and
  Ploog}}]{ramsteiner02}
\bibinfo{author}{\bibfnamefont{M.}~\bibnamefont{Ramsteiner}},
  \bibinfo{author}{\bibfnamefont{H.}~\bibnamefont{Hao}},
  \bibinfo{author}{\bibfnamefont{A.}~\bibnamefont{Kawaharazuka}},
  \bibinfo{author}{\bibfnamefont{H.}~\bibnamefont{Zhu}},
  \bibinfo{author}{\bibfnamefont{M.}~\bibnamefont{K{\"a}stner}},
  \bibinfo{author}{\bibfnamefont{R.}~\bibnamefont{Hey}},
  \bibinfo{author}{\bibfnamefont{L.}~\bibnamefont{D{\"a}weritz}},
  \bibinfo{author}{\bibfnamefont{H.}~\bibnamefont{Grahn}}, \bibnamefont{and}
  \bibinfo{author}{\bibfnamefont{K.}~\bibnamefont{Ploog}},
  \bibinfo{journal}{Phys.\ Rev.} \textbf{\bibinfo{volume}{B66}},
  \bibinfo{pages}{081304} (\bibinfo{year}{2002}).

\bibitem[{\citenamefont{Ney et~al.}(2004)\citenamefont{Ney, Hasjedal, Pampuch,
  Das, D{\"a}weritz, Koch, Ploog, Toli{\'n}ski, Lindner, Lenz et~al.}}]{ney04}
\bibinfo{author}{\bibfnamefont{A.}~\bibnamefont{Ney}},
  \bibinfo{author}{\bibfnamefont{T.}~\bibnamefont{Hasjedal}},
  \bibinfo{author}{\bibfnamefont{C.}~\bibnamefont{Pampuch}},
  \bibinfo{author}{\bibfnamefont{A.}~\bibnamefont{Das}},
  \bibinfo{author}{\bibfnamefont{L.}~\bibnamefont{D{\"a}weritz}},
  \bibinfo{author}{\bibfnamefont{R.}~\bibnamefont{Koch}},
  \bibinfo{author}{\bibfnamefont{K.}~\bibnamefont{Ploog}},
  \bibinfo{author}{\bibfnamefont{T.}~\bibnamefont{Toli{\'n}ski}},
  \bibinfo{author}{\bibfnamefont{J.}~\bibnamefont{Lindner}},
  \bibinfo{author}{\bibfnamefont{K.}~\bibnamefont{Lenz}}, \bibnamefont{et~al.},
  \bibinfo{journal}{Phys.\ Rev.} \textbf{\bibinfo{volume}{B 69}},
  \bibinfo{pages}{081306(R)} (\bibinfo{year}{2004}), \bibinfo{note}{and
  references therein}.

\bibitem[{\citenamefont{Tanaka et~al.}(1994)\citenamefont{Tanaka, Harbison,
  Park, Park, Shin, and Rothberg}}]{tanaka94}
\bibinfo{author}{\bibfnamefont{M.}~\bibnamefont{Tanaka}},
  \bibinfo{author}{\bibfnamefont{J.}~\bibnamefont{Harbison}},
  \bibinfo{author}{\bibfnamefont{M.}~\bibnamefont{Park}},
  \bibinfo{author}{\bibfnamefont{Y.}~\bibnamefont{Park}},
  \bibinfo{author}{\bibfnamefont{T.}~\bibnamefont{Shin}}, \bibnamefont{and}
  \bibinfo{author}{\bibfnamefont{G.}~\bibnamefont{Rothberg}},
  \bibinfo{journal}{Appl.\ Phys.\ Lett.} \textbf{\bibinfo{volume}{65}},
  \bibinfo{pages}{1964} (\bibinfo{year}{1994}).

\bibitem[{\citenamefont{Sadowski
  et~al.}(2000{\natexlab{a}})\citenamefont{Sadowski, Kanski, Ilver, and
  Johansson}}]{sadowski00}
\bibinfo{author}{\bibfnamefont{J.}~\bibnamefont{Sadowski}},
  \bibinfo{author}{\bibfnamefont{J.}~\bibnamefont{Kanski}},
  \bibinfo{author}{\bibfnamefont{L.}~\bibnamefont{Ilver}}, \bibnamefont{and}
  \bibinfo{author}{\bibfnamefont{J.}~\bibnamefont{Johansson}},
  \bibinfo{journal}{Appl.\ Surf.\ Sci.} \textbf{\bibinfo{volume}{166}},
  \bibinfo{pages}{247} (\bibinfo{year}{2000}{\natexlab{a}}).

\bibitem[{\citenamefont{Sadowski et~al.}(2005)\citenamefont{Sadowski, Adell,
  Kanski, Ilver, Janik, {\L}usakowska, Domaga{\l}a, Kret, D{\l}u{\.z}ewski,
  Brucas et~al.}}]{sadowski05apl}
\bibinfo{author}{\bibfnamefont{J.}~\bibnamefont{Sadowski}},
  \bibinfo{author}{\bibfnamefont{M.}~\bibnamefont{Adell}},
  \bibinfo{author}{\bibfnamefont{J.}~\bibnamefont{Kanski}},
  \bibinfo{author}{\bibfnamefont{L.}~\bibnamefont{Ilver}},
  \bibinfo{author}{\bibfnamefont{E.}~\bibnamefont{Janik}},
  \bibinfo{author}{\bibfnamefont{E.}~\bibnamefont{{\L}usakowska}},
  \bibinfo{author}{\bibfnamefont{J.}~\bibnamefont{Domaga{\l}a}},
  \bibinfo{author}{\bibfnamefont{S.}~\bibnamefont{Kret}},
  \bibinfo{author}{\bibfnamefont{P.}~\bibnamefont{D{\l}u{\.z}ewski}},
  \bibinfo{author}{\bibfnamefont{R.}~\bibnamefont{Brucas}},
  \bibnamefont{et~al.}, \bibinfo{journal}{Appl.\ Phys.\ Lett.}
  \textbf{\bibinfo{volume}{87}}, \bibinfo{pages}{263114}
  (\bibinfo{year}{2005}).

\bibitem[{\citenamefont{Berry et~al.}(2000)\citenamefont{Berry, Chun, Ku,
  Samarth, Malajovich, and Awschalom}}]{berry00}
\bibinfo{author}{\bibfnamefont{J.}~\bibnamefont{Berry}},
  \bibinfo{author}{\bibfnamefont{S.}~\bibnamefont{Chun}},
  \bibinfo{author}{\bibfnamefont{K.}~\bibnamefont{Ku}},
  \bibinfo{author}{\bibfnamefont{N.}~\bibnamefont{Samarth}},
  \bibinfo{author}{\bibfnamefont{I.}~\bibnamefont{Malajovich}},
  \bibnamefont{and}
  \bibinfo{author}{\bibfnamefont{D.}~\bibnamefont{Awschalom}},
  \bibinfo{journal}{Appl.\ Phys.\ Lett.} \textbf{\bibinfo{volume}{77}},
  \bibinfo{pages}{3812} (\bibinfo{year}{2000}).

\bibitem[{\citenamefont{Ploog}(2001)}]{ploog01}
\bibinfo{author}{\bibfnamefont{K.}~\bibnamefont{Ploog}},
  \bibinfo{journal}{Mat.\ Sci.\ Semicond.\ Process.}
  \textbf{\bibinfo{volume}{4}}, \bibinfo{pages}{451} (\bibinfo{year}{2001}).

\bibitem[{\citenamefont{Trampert}(2002)}]{trampert02}
\bibinfo{author}{\bibfnamefont{A.}~\bibnamefont{Trampert}},
  \bibinfo{journal}{Physica E} \textbf{\bibinfo{volume}{13}},
  \bibinfo{pages}{1119} (\bibinfo{year}{2002}).

\bibitem[{\citenamefont{Kowalski
  et~al.}(2004{\natexlab{a}})\citenamefont{Kowalski, Kowalik, Iwanowski,
  Lusakowska, Sawicki, Sadowski, Grzegory, and Porowski}}]{kowalski04app}
\bibinfo{author}{\bibfnamefont{B.}~\bibnamefont{Kowalski}},
  \bibinfo{author}{\bibfnamefont{I.}~\bibnamefont{Kowalik}},
  \bibinfo{author}{\bibfnamefont{R.}~\bibnamefont{Iwanowski}},
  \bibinfo{author}{\bibfnamefont{E.}~\bibnamefont{Lusakowska}},
  \bibinfo{author}{\bibfnamefont{M.}~\bibnamefont{Sawicki}},
  \bibinfo{author}{\bibfnamefont{J.}~\bibnamefont{Sadowski}},
  \bibinfo{author}{\bibfnamefont{I.}~\bibnamefont{Grzegory}}, \bibnamefont{and}
  \bibinfo{author}{\bibfnamefont{S.}~\bibnamefont{Porowski}},
  \bibinfo{journal}{Acta Phys.\ Pol.} \textbf{\bibinfo{volume}{105}},
  \bibinfo{pages}{645} (\bibinfo{year}{2004}{\natexlab{a}}).

\bibitem[{\citenamefont{Kowalski
  et~al.}(2004{\natexlab{b}})\citenamefont{Kowalski, Iwanowski, Sadowski,
  Kowalik, Kanski, Grzegory, and Porowski}}]{kowalski04ss}
\bibinfo{author}{\bibfnamefont{B.}~\bibnamefont{Kowalski}},
  \bibinfo{author}{\bibfnamefont{R.}~\bibnamefont{Iwanowski}},
  \bibinfo{author}{\bibfnamefont{J.}~\bibnamefont{Sadowski}},
  \bibinfo{author}{\bibfnamefont{I.}~\bibnamefont{Kowalik}},
  \bibinfo{author}{\bibfnamefont{J.}~\bibnamefont{Kanski}},
  \bibinfo{author}{\bibfnamefont{I.}~\bibnamefont{Grzegory}}, \bibnamefont{and}
  \bibinfo{author}{\bibfnamefont{S.}~\bibnamefont{Porowski}},
  \bibinfo{journal}{Surf.\ Sci.} \textbf{\bibinfo{volume}{548}},
  \bibinfo{pages}{220} (\bibinfo{year}{2004}{\natexlab{b}}).

\bibitem[{\citenamefont{Sadowski
  et~al.}(2000{\natexlab{b}})\citenamefont{Sadowski, Domaga{\l}a,
  B\c{a}k-Misiuk, Kolesnik, Sawicki, \'Swiatek, Kanski, Ilver, and
  Str{\"o}m}}]{sadowski00jvst}
\bibinfo{author}{\bibfnamefont{J.}~\bibnamefont{Sadowski}},
  \bibinfo{author}{\bibfnamefont{J.~Z.} \bibnamefont{Domaga{\l}a}},
  \bibinfo{author}{\bibfnamefont{J.}~\bibnamefont{B\c{a}k-Misiuk}},
  \bibinfo{author}{\bibfnamefont{S.}~\bibnamefont{Kolesnik}},
  \bibinfo{author}{\bibfnamefont{M.}~\bibnamefont{Sawicki}},
  \bibinfo{author}{\bibfnamefont{K.}~\bibnamefont{\'Swiatek}},
  \bibinfo{author}{\bibfnamefont{J.}~\bibnamefont{Kanski}},
  \bibinfo{author}{\bibfnamefont{L.}~\bibnamefont{Ilver}}, \bibnamefont{and}
  \bibinfo{author}{\bibfnamefont{V.}~\bibnamefont{Str{\"o}m}},
  \bibinfo{journal}{J.\ Vac.\ Sci.\ Technol.\ B} \textbf{\bibinfo{volume}{18}},
  \bibinfo{pages}{1697} (\bibinfo{year}{2000}{\natexlab{b}}).

\bibitem[{\citenamefont{{\AA}sklund et~al.}(2001)\citenamefont{{\AA}sklund,
  Ilver, Kanski, Mankefors, S{\"o}dervall, and Sadowski}}]{asklund01}
\bibinfo{author}{\bibfnamefont{H.}~\bibnamefont{{\AA}sklund}},
  \bibinfo{author}{\bibfnamefont{L.}~\bibnamefont{Ilver}},
  \bibinfo{author}{\bibfnamefont{J.}~\bibnamefont{Kanski}},
  \bibinfo{author}{\bibfnamefont{S.}~\bibnamefont{Mankefors}},
  \bibinfo{author}{\bibfnamefont{U.}~\bibnamefont{S{\"o}dervall}},
  \bibnamefont{and} \bibinfo{author}{\bibfnamefont{J.}~\bibnamefont{Sadowski}},
  \bibinfo{journal}{Phys.\ Rev.\ B} \textbf{\bibinfo{volume}{63}},
  \bibinfo{pages}{195314} (\bibinfo{year}{2001}), \bibinfo{note}{and references
  therein}.

\bibitem[{\citenamefont{Okabayashi et~al.}(2004)\citenamefont{Okabayashi,
  Mizuguchi, Ono, Oshima, Fujimori, Kuramochi, and Akinaga}}]{okabayashi04}
\bibinfo{author}{\bibfnamefont{J.}~\bibnamefont{Okabayashi}},
  \bibinfo{author}{\bibfnamefont{M.}~\bibnamefont{Mizuguchi}},
  \bibinfo{author}{\bibfnamefont{K.}~\bibnamefont{Ono}},
  \bibinfo{author}{\bibfnamefont{M.}~\bibnamefont{Oshima}},
  \bibinfo{author}{\bibfnamefont{A.}~\bibnamefont{Fujimori}},
  \bibinfo{author}{\bibfnamefont{H.}~\bibnamefont{Kuramochi}},
  \bibnamefont{and} \bibinfo{author}{\bibfnamefont{H.}~\bibnamefont{Akinaga}},
  \bibinfo{journal}{Phys.\ Rev. B} \textbf{\bibinfo{volume}{70}},
  \bibinfo{pages}{233305} (\bibinfo{year}{2004}).

\bibitem[{\citenamefont{Moreno et~al.}(2002)\citenamefont{Moreno, Trampert,
  Jenichen, D{\"a}weritz, and Ploog}}]{moreno02}
\bibinfo{author}{\bibfnamefont{M.}~\bibnamefont{Moreno}},
  \bibinfo{author}{\bibfnamefont{A.}~\bibnamefont{Trampert}},
  \bibinfo{author}{\bibfnamefont{B.}~\bibnamefont{Jenichen}},
  \bibinfo{author}{\bibfnamefont{L.}~\bibnamefont{D{\"a}weritz}},
  \bibnamefont{and} \bibinfo{author}{\bibfnamefont{K.}~\bibnamefont{Ploog}},
  \bibinfo{journal}{J.\ Appl.\ Phys.} \textbf{\bibinfo{volume}{92}},
  \bibinfo{pages}{4672} (\bibinfo{year}{2002}).

\bibitem[{\citenamefont{Fano}(1961)}]{fano61}
\bibinfo{author}{\bibfnamefont{U.}~\bibnamefont{Fano}},
  \bibinfo{journal}{Phys.\ Rev.} \textbf{\bibinfo{volume}{124}},
  \bibinfo{pages}{1866} (\bibinfo{year}{1961}).

\bibitem[{\citenamefont{Kowalik et~al.}(2004)\citenamefont{Kowalik, Kowalski,
  Orlowski, Lusakowska, Iwanowski, Mickevicius, Johnson, Grzegory, and
  Porowski}}]{kowalik04}
\bibinfo{author}{\bibfnamefont{I.}~\bibnamefont{Kowalik}},
  \bibinfo{author}{\bibfnamefont{B.}~\bibnamefont{Kowalski}},
  \bibinfo{author}{\bibfnamefont{B.}~\bibnamefont{Orlowski}},
  \bibinfo{author}{\bibfnamefont{E.}~\bibnamefont{Lusakowska}},
  \bibinfo{author}{\bibfnamefont{R.}~\bibnamefont{Iwanowski}},
  \bibinfo{author}{\bibfnamefont{S.}~\bibnamefont{Mickevicius}},
  \bibinfo{author}{\bibfnamefont{R.}~\bibnamefont{Johnson}},
  \bibinfo{author}{\bibfnamefont{I.}~\bibnamefont{Grzegory}}, \bibnamefont{and}
  \bibinfo{author}{\bibfnamefont{S.}~\bibnamefont{Porowski}},
  \bibinfo{journal}{Surf.\ Sci.} \textbf{\bibinfo{volume}{566-568}},
  \bibinfo{pages}{457} (\bibinfo{year}{2004}).

\bibitem[{\citenamefont{Shimada et~al.}(98)\citenamefont{Shimada, Rader,
  Fujimori, Kimura, Ono, Kamakura, Kakizaki, Tanaka, and Shirai}}]{shimada98}
\bibinfo{author}{\bibfnamefont{K.}~\bibnamefont{Shimada}},
  \bibinfo{author}{\bibfnamefont{O.}~\bibnamefont{Rader}},
  \bibinfo{author}{\bibfnamefont{A.}~\bibnamefont{Fujimori}},
  \bibinfo{author}{\bibfnamefont{A.}~\bibnamefont{Kimura}},
  \bibinfo{author}{\bibfnamefont{K.}~\bibnamefont{Ono}},
  \bibinfo{author}{\bibfnamefont{N.}~\bibnamefont{Kamakura}},
  \bibinfo{author}{\bibfnamefont{A.}~\bibnamefont{Kakizaki}},
  \bibinfo{author}{\bibfnamefont{M.}~\bibnamefont{Tanaka}}, \bibnamefont{and}
  \bibinfo{author}{\bibfnamefont{M.}~\bibnamefont{Shirai}},
  \bibinfo{journal}{J. Electron Spectr.\ Rel.\ Phenom.}
  \textbf{\bibinfo{volume}{88-91}}, \bibinfo{pages}{207} (\bibinfo{year}{98}).

\bibitem[{\citenamefont{Shimada et~al.}(1999)\citenamefont{Shimada, Rader,
  Fujimori, Kimura, Ono, Kamakura, Kakizaki, Tanaka, and Shirai}}]{shimada99}
\bibinfo{author}{\bibfnamefont{K.}~\bibnamefont{Shimada}},
  \bibinfo{author}{\bibfnamefont{O.}~\bibnamefont{Rader}},
  \bibinfo{author}{\bibfnamefont{A.}~\bibnamefont{Fujimori}},
  \bibinfo{author}{\bibfnamefont{A.}~\bibnamefont{Kimura}},
  \bibinfo{author}{\bibfnamefont{K.}~\bibnamefont{Ono}},
  \bibinfo{author}{\bibfnamefont{N.}~\bibnamefont{Kamakura}},
  \bibinfo{author}{\bibfnamefont{A.}~\bibnamefont{Kakizaki}},
  \bibinfo{author}{\bibfnamefont{M.}~\bibnamefont{Tanaka}}, \bibnamefont{and}
  \bibinfo{author}{\bibfnamefont{M.}~\bibnamefont{Shirai}},
  \bibinfo{journal}{J. Electron Spectr.\ Rel.\ Phenom.}
  \textbf{\bibinfo{volume}{101}}, \bibinfo{pages}{383} (\bibinfo{year}{1999}).

\bibitem[{\citenamefont{Ono et~al.}(2002)\citenamefont{Ono, Okabayashi,
  Mizuguchi, Oshima, Fujimori, and Akinaga}}]{ono02}
\bibinfo{author}{\bibfnamefont{K.}~\bibnamefont{Ono}},
  \bibinfo{author}{\bibfnamefont{J.}~\bibnamefont{Okabayashi}},
  \bibinfo{author}{\bibfnamefont{M.}~\bibnamefont{Mizuguchi}},
  \bibinfo{author}{\bibfnamefont{M.}~\bibnamefont{Oshima}},
  \bibinfo{author}{\bibfnamefont{A.}~\bibnamefont{Fujimori}}, \bibnamefont{and}
  \bibinfo{author}{\bibfnamefont{H.}~\bibnamefont{Akinaga}},
  \bibinfo{journal}{J.\ Appl.\ Phys.} \textbf{\bibinfo{volume}{91}},
  \bibinfo{pages}{8088} (\bibinfo{year}{2002}).

\bibitem[{\citenamefont{Kim et~al.}(2006)\citenamefont{Kim, Jeon, Kang, Lee,
  Lee, and Jin}}]{kim06}
\bibinfo{author}{\bibfnamefont{T.}~\bibnamefont{Kim}},
  \bibinfo{author}{\bibfnamefont{H.}~\bibnamefont{Jeon}},
  \bibinfo{author}{\bibfnamefont{T.}~\bibnamefont{Kang}},
  \bibinfo{author}{\bibfnamefont{H.}~\bibnamefont{Lee}},
  \bibinfo{author}{\bibfnamefont{J.}~\bibnamefont{Lee}}, \bibnamefont{and}
  \bibinfo{author}{\bibfnamefont{S.}~\bibnamefont{Jin}},
  \bibinfo{journal}{Appl.\ Phys.\ Lett.} \textbf{\bibinfo{volume}{88}},
  \bibinfo{pages}{021915} (\bibinfo{year}{2006}).

\bibitem[{\citenamefont{Paszkowicz}()}]{paszkowicz05}
\bibinfo{author}{\bibfnamefont{W.}~\bibnamefont{Paszkowicz}},
  \bibinfo{note}{private communication}.

\bibitem[{\citenamefont{Shirai et~al.}(1998)\citenamefont{Shirai, Ogawa,
  Kitagawa, and Suzuki}}]{shirai98jmm}
\bibinfo{author}{\bibfnamefont{M.}~\bibnamefont{Shirai}},
  \bibinfo{author}{\bibfnamefont{T.}~\bibnamefont{Ogawa}},
  \bibinfo{author}{\bibfnamefont{I.}~\bibnamefont{Kitagawa}}, \bibnamefont{and}
  \bibinfo{author}{\bibfnamefont{N.}~\bibnamefont{Suzuki}},
  \bibinfo{journal}{J.\ Magn.\ Magn.\ Mat.} \textbf{\bibinfo{volume}{177-181}},
  \bibinfo{pages}{1383} (\bibinfo{year}{1998}).

\bibitem[{\citenamefont{Zhao et~al.}(2002)\citenamefont{Zhao, Geng, Freeman,
  and Delley}}]{zhao02}
\bibinfo{author}{\bibfnamefont{Y.-J.} \bibnamefont{Zhao}},
  \bibinfo{author}{\bibfnamefont{W.}~\bibnamefont{Geng}},
  \bibinfo{author}{\bibfnamefont{A.}~\bibnamefont{Freeman}}, \bibnamefont{and}
  \bibinfo{author}{\bibfnamefont{B.}~\bibnamefont{Delley}},
  \bibinfo{journal}{Phys.\ Rev.\ B} \textbf{\bibinfo{volume}{65}},
  \bibinfo{pages}{113202} (\bibinfo{year}{2002}).

\bibitem[{\citenamefont{Kim and Freeman}(2004)}]{kim04}
\bibinfo{author}{\bibfnamefont{M.}~\bibnamefont{Kim}} \bibnamefont{and}
  \bibinfo{author}{\bibfnamefont{A.}~\bibnamefont{Freeman}},
  \bibinfo{journal}{Appl.\ Phys.\ Lett.} \textbf{\bibinfo{volume}{85}},
  \bibinfo{pages}{4983} (\bibinfo{year}{2004}).

\bibitem[{\citenamefont{Hong et~al.}(2005)\citenamefont{Hong, Kang, Kang, Lee,
  Kim, and Chang}}]{hong05}
\bibinfo{author}{\bibfnamefont{H.-M.} \bibnamefont{Hong}},
  \bibinfo{author}{\bibfnamefont{Y.-J.} \bibnamefont{Kang}},
  \bibinfo{author}{\bibfnamefont{J.}~\bibnamefont{Kang}},
  \bibinfo{author}{\bibfnamefont{E.-C.} \bibnamefont{Lee}},
  \bibinfo{author}{\bibfnamefont{Y.-H.} \bibnamefont{Kim}}, \bibnamefont{and}
  \bibinfo{author}{\bibfnamefont{K.}~\bibnamefont{Chang}},
  \bibinfo{journal}{Phys.\ Rev.\ B} \textbf{\bibinfo{volume}{72}},
  \bibinfo{pages}{144408} (\bibinfo{year}{2005}).

\bibitem[{\citenamefont{Zheng and Davenport}(2004)}]{zheng04}
\bibinfo{author}{\bibfnamefont{J.-C.} \bibnamefont{Zheng}} \bibnamefont{and}
  \bibinfo{author}{\bibfnamefont{J.}~\bibnamefont{Davenport}},
  \bibinfo{journal}{Phys.\ Rev.\ B} \textbf{\bibinfo{volume}{69}},
  \bibinfo{pages}{144415} (\bibinfo{year}{2004}).

\bibitem[{\citenamefont{Debernardi
  et~al.}(2003{\natexlab{a}})\citenamefont{Debernardi, Peressi, and
  Baldereschi}}]{debernardi03cms}
\bibinfo{author}{\bibfnamefont{A.}~\bibnamefont{Debernardi}},
  \bibinfo{author}{\bibfnamefont{M.}~\bibnamefont{Peressi}}, \bibnamefont{and}
  \bibinfo{author}{\bibfnamefont{A.}~\bibnamefont{Baldereschi}},
  \bibinfo{journal}{Comput.\ Mater.\ Sci.} \textbf{\bibinfo{volume}{27}},
  \bibinfo{pages}{175} (\bibinfo{year}{2003}{\natexlab{a}}).

\bibitem[{\citenamefont{Debernardi
  et~al.}(2003{\natexlab{b}})\citenamefont{Debernardi, Peressi, and
  Baldereschi}}]{debernardi03mse}
\bibinfo{author}{\bibfnamefont{A.}~\bibnamefont{Debernardi}},
  \bibinfo{author}{\bibfnamefont{M.}~\bibnamefont{Peressi}}, \bibnamefont{and}
  \bibinfo{author}{\bibfnamefont{A.}~\bibnamefont{Baldereschi}},
  \bibinfo{journal}{Mater.\ Sci.\ Eng.\ C} \textbf{\bibinfo{volume}{23}},
  \bibinfo{pages}{1059} (\bibinfo{year}{2003}{\natexlab{b}}).

\bibitem[{\citenamefont{Kitamura et~al.}(2006)\citenamefont{Kitamura, Shen,
  Sugiyama, Nakanishi, Shimizu, and Okumura}}]{kitamura06}
\bibinfo{author}{\bibfnamefont{T.}~\bibnamefont{Kitamura}},
  \bibinfo{author}{\bibfnamefont{X.-Q.} \bibnamefont{Shen}},
  \bibinfo{author}{\bibfnamefont{M.}~\bibnamefont{Sugiyama}},
  \bibinfo{author}{\bibfnamefont{H.}~\bibnamefont{Nakanishi}},
  \bibinfo{author}{\bibfnamefont{M.}~\bibnamefont{Shimizu}}, \bibnamefont{and}
  \bibinfo{author}{\bibfnamefont{H.}~\bibnamefont{Okumura}},
  \bibinfo{journal}{Jpn.\ J.\ Appl.\ Phys.} \textbf{\bibinfo{volume}{45}},
  \bibinfo{pages}{57} (\bibinfo{year}{2006}).

\bibitem[{\citenamefont{Eastman et~al.}(1980)\citenamefont{Eastman, Chiang,
  Heimann, and Himpsel}}]{eastman80}
\bibinfo{author}{\bibfnamefont{D.}~\bibnamefont{Eastman}},
  \bibinfo{author}{\bibfnamefont{T.-C.} \bibnamefont{Chiang}},
  \bibinfo{author}{\bibfnamefont{P.}~\bibnamefont{Heimann}}, \bibnamefont{and}
  \bibinfo{author}{\bibfnamefont{F.}~\bibnamefont{Himpsel}},
  \bibinfo{journal}{Phys.\ Rev.\ Lett.} \textbf{\bibinfo{volume}{45}},
  \bibinfo{pages}{656} (\bibinfo{year}{1980}).

\bibitem[{\citenamefont{Prietsch et~al.}(1988)\citenamefont{Prietsch,
  Laubschat, Domke, and Kaindl}}]{prietsch88}
\bibinfo{author}{\bibfnamefont{M.}~\bibnamefont{Prietsch}},
  \bibinfo{author}{\bibfnamefont{C.}~\bibnamefont{Laubschat}},
  \bibinfo{author}{\bibfnamefont{M.}~\bibnamefont{Domke}}, \bibnamefont{and}
  \bibinfo{author}{\bibfnamefont{G.}~\bibnamefont{Kaindl}},
  \bibinfo{journal}{Phys.\ Rev.\ B} \textbf{\bibinfo{volume}{38}},
  \bibinfo{pages}{10655} (\bibinfo{year}{1988}).

\bibitem[{\citenamefont{Chaika et~al.}(1999)\citenamefont{Chaika, Grazhulis,
  Ionov, Molodtsov, and Laubschat}}]{chaika99}
\bibinfo{author}{\bibfnamefont{A.}~\bibnamefont{Chaika}},
  \bibinfo{author}{\bibfnamefont{V.}~\bibnamefont{Grazhulis}},
  \bibinfo{author}{\bibfnamefont{A.}~\bibnamefont{Ionov}},
  \bibinfo{author}{\bibfnamefont{S.}~\bibnamefont{Molodtsov}},
  \bibnamefont{and}
  \bibinfo{author}{\bibfnamefont{C.}~\bibnamefont{Laubschat}},
  \bibinfo{journal}{Appl.\ Surf.\ Sci.} \textbf{\bibinfo{volume}{143}},
  \bibinfo{pages}{277} (\bibinfo{year}{1999}).

\bibitem[{\citenamefont{Barth et~al.}(1979)\citenamefont{Barth, Kalkoffen, and
  Kunz}}]{barth79}
\bibinfo{author}{\bibfnamefont{J.}~\bibnamefont{Barth}},
  \bibinfo{author}{\bibfnamefont{G.}~\bibnamefont{Kalkoffen}},
  \bibnamefont{and} \bibinfo{author}{\bibfnamefont{C.}~\bibnamefont{Kunz}},
  \bibinfo{journal}{Phys.\ Lett.} \textbf{\bibinfo{volume}{74A}},
  \bibinfo{pages}{360} (\bibinfo{year}{1979}).

\bibitem[{\citenamefont{Kakizaki et~al.}(1983)\citenamefont{Kakizaki, Sugeno,
  Ishii, Sugawara, Nagakura, and Shin}}]{kakizaki83}
\bibinfo{author}{\bibfnamefont{A.}~\bibnamefont{Kakizaki}},
  \bibinfo{author}{\bibfnamefont{K.}~\bibnamefont{Sugeno}},
  \bibinfo{author}{\bibfnamefont{T.}~\bibnamefont{Ishii}},
  \bibinfo{author}{\bibfnamefont{H.}~\bibnamefont{Sugawara}},
  \bibinfo{author}{\bibfnamefont{I.}~\bibnamefont{Nagakura}}, \bibnamefont{and}
  \bibinfo{author}{\bibfnamefont{S.}~\bibnamefont{Shin}},
  \bibinfo{journal}{Phys.\ Rev.\ B} \textbf{\bibinfo{volume}{28}},
  \bibinfo{pages}{1026} (\bibinfo{year}{1983}).

\bibitem[{\citenamefont{Ley et~al.}(1987)\citenamefont{Ley, Taniguchi, Ghijsen,
  and Johnson}}]{ley87}
\bibinfo{author}{\bibfnamefont{L.}~\bibnamefont{Ley}},
  \bibinfo{author}{\bibfnamefont{M.}~\bibnamefont{Taniguchi}},
  \bibinfo{author}{\bibfnamefont{J.}~\bibnamefont{Ghijsen}}, \bibnamefont{and}
  \bibinfo{author}{\bibfnamefont{R.}~\bibnamefont{Johnson}},
  \bibinfo{journal}{Phys.\ Rev.\ B} \textbf{\bibinfo{volume}{35}},
  \bibinfo{pages}{2839} (\bibinfo{year}{1987}).

\bibitem[{\citenamefont{Davis}(1986)}]{davis86}
\bibinfo{author}{\bibfnamefont{L.}~\bibnamefont{Davis}}, \bibinfo{journal}{J.\
  Appl.\ Phys.} \textbf{\bibinfo{volume}{59}}, \bibinfo{pages}{R25}
  (\bibinfo{year}{1986}), \bibinfo{note}{and references therein}.

\bibitem[{\citenamefont{Ploog et~al.}(2006)\citenamefont{Ploog, D{\"a}weritz,
  Engel-Herbert, and Hesjedal}}]{ploog06}
\bibinfo{author}{\bibfnamefont{K.}~\bibnamefont{Ploog}},
  \bibinfo{author}{\bibfnamefont{L.}~\bibnamefont{D{\"a}weritz}},
  \bibinfo{author}{\bibfnamefont{R.}~\bibnamefont{Engel-Herbert}},
  \bibnamefont{and} \bibinfo{author}{\bibfnamefont{T.}~\bibnamefont{Hesjedal}},
  \bibinfo{journal}{phys.\ stat.\ sol. (a)} \textbf{\bibinfo{volume}{203}},
  \bibinfo{pages}{3574} (\bibinfo{year}{2006}).

\end{thebibliography}

\end{document}